\documentclass[10pt,logo]{uwtechreport}
\usepackage[round,authoryear]{natbib}
\usepackage{flafter}

\usepackage{latexsym}
\usepackage{subcaption}
\usepackage{array}
\usepackage{multirow}
\usepackage{tikz}
\usetikzlibrary{arrows.meta,positioning,calc,fit,backgrounds}
\definecolor{stepA}{RGB}{225,238,250}
\definecolor{stepB}{RGB}{252,239,224}
\definecolor{stepC}{RGB}{228,244,232}
\definecolor{stepD}{RGB}{243,238,250}
\definecolor{inkgray}{RGB}{90,90,90}
\newcolumntype{L}[1]{>{\raggedright\arraybackslash}p{#1}}
\DeclareRobustCommand{\yes}{\tikz[baseline=-0.5ex]{\filldraw[line width=0.4pt] (0,0) circle (0.32em);}}
\DeclareRobustCommand{\no}{\tikz[baseline=-0.5ex]{\draw[line width=0.4pt] (0,0) circle (0.32em);}}
\DeclareRobustCommand{\partly}{\tikz[baseline=-0.5ex]{\fill (0,0.32em) arc[start angle=90,end angle=270,radius=0.32em] -- cycle;\draw[line width=0.4pt] (0,0) circle (0.32em);}}
\newcommand{\ourwork}{\texttt{WebUniverse}}
\newcommand{\ourdata}{\ourwork{}\texttt{-Mix}}
\newcommand{\ourenv}{\ourwork{}\texttt{-Env}}
\newcommand{\ourmodel}{Qwen3.5-WebUniverse}
\newcommand{\modelicon}[1]{\raisebox{-.12em}{#1}\hspace{.3em}}

\usepackage{placeins}


\usepackage{fvextra}
\usepackage{newunicodechar}
\newunicodechar{’}{\textquoteright}
\newunicodechar{“}{\textquotedblleft}
\newunicodechar{”}{\textquotedblright}
\newunicodechar{→}{\ensuremath{\rightarrow}}
\newunicodechar{✓}{\ensuremath{\checkmark}}
\usepackage[breakable,skins]{tcolorbox}

\title{\ourwork{}: Scaling Verified Training Data for Agentic Web Development}
\author{
  Yaorui Ding$^{1\dagger}$ \; Fengqing Jiang$^{1\dagger\ddagger}$ \; Yichen Feng$^{1\dagger}$ \; Yuetai Li$^{1}$ \; Zhangchen Xu$^{1}$ \; Kaiyuan Zheng$^{1}$ \; \newline
  Xuanyu Lu$^{1}$ \; Bhaskar Ramasubramanian$^{1,2}$ \; Basel Alomair$^{1,3,4}$ \; Luyao Niu$^{1}$ \; Radha Poovendran$^{1}$ \\
  \small $^{1}$University of Washington \quad $^{2}$Western Washington University \\
  \small $^{3}$King Abdulaziz City for Science and Technology \quad $^{4}$HUMAIN \\
  \small $^{\dagger}$Equal Contribution \quad $^{\ddagger}$Project Lead \\
  {\small \href{https://webuniverse-nsl.github.io/}{\texttt{webuniverse-nsl.github.io}}}
}

\begin{document}
\maketitle
\begin{abstract}
Training agents for front-end web development requires demonstrations of code editing, execution, browser inspection and revision, yet public trajectories and reusable task environments remain limited.
We introduce \ourwork{}, a pipeline that constructs verifiable tasks with verified development trajectories from open-source repositories and real web pages.
The tasks cover text-grounded application development in standalone and iterative settings and vision-grounded reproduction across pages and viewports.
We collect agent trajectories for these tasks and verify the resulting applications against task-specific functional or visual requirements.
We release \ourenv{}, containing 15K executable task environments, and \ourdata{}, containing the corresponding 15K development trajectories with verification records and scores. Fine-tuning Qwen3.5-35B-A3B on \ourdata{} raises the Vision2Web average score from 51.7 to 68.3, WebGen-Bench accuracy from 34.5\% to 63.9\%, and Web-Bench pass@2 from 12.2\% to 23.1\%.
The resulting model also outperforms a public-data baseline trained with the same configuration and sample budget across all reported metrics.
Ablations show complementary benefits from text and visual trajectories and continued gains as training data scales from 5K to 15K trajectories.
\end{abstract}

\abscontent

\par\vspace{10pt}
\noindent\begin{minipage}{\textwidth}
\centering
\includegraphics[width=\textwidth]{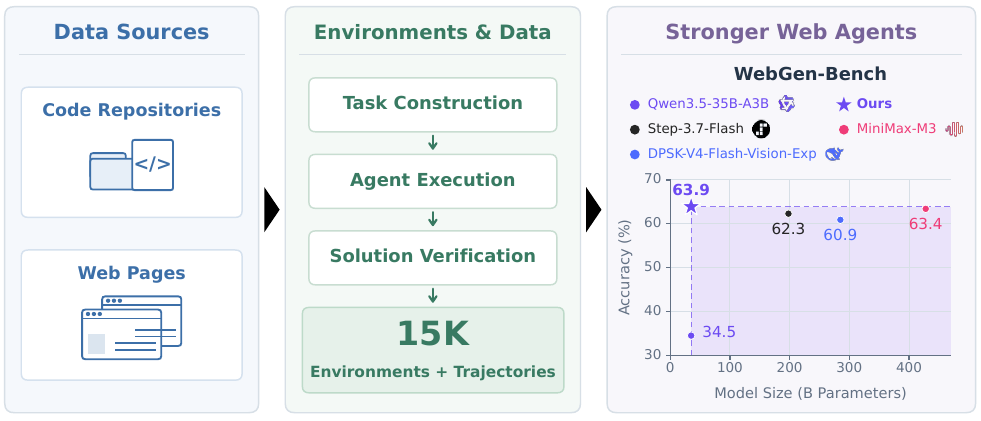}
\captionof{figure}{\textbf{Left:} Data sources for \ourwork{}. \textbf{Middle:} Pipeline for synthesizing environments and trajectories. \textbf{Right:} Fine-tuning Qwen3.5-35B-A3B on \ourdata{} raises WebGen-Bench accuracy from 34.5\% to 63.9\%, outperforming leading open-source models with substantially larger parameter counts. See Sections~\ref{sec:construction} and~\ref{sec:experiments} for details.}
\label{fig:teaser}
\end{minipage}%
\clearpage

\section{Introduction}

\begin{table}[t]
\centering
\setlength{\tabcolsep}{3pt}
\small
\renewcommand{\arraystretch}{1.12}
\resizebox{\textwidth}{!}{%
\begin{tabular}{lcccccccc}
\toprule
     & Resource & Front-end & Visual & Iterative & Runtime & Public & Public & Verif. \\
     & forms & dev. & refs. & dev. & checks & envs. & data & records \\
\midrule
SWE-Flow~\citep{sweflow} & Pair+ENV & \no & \no & \yes & \yes & \partly & \no & \no \\
CLI-Universe~\citep{cliuniverse} & AT+ENV & \no & \no & \no & \yes & \no & \no & \no \\
CodeMidas~\citep{codemidas} & ENV & \no & \no & \no & \yes & \no & \no & \no \\
\midrule
WebSight~\citep{websight} & Pair & \yes & \yes & \no & \no & \no & \yes & \no \\
WebGen-Bench (train)~\citep{webgenbench} & Pair+ENV & \yes & \no & \no & \yes & \yes & \yes & \no \\
WebGen-Agent~\citep{webgenagent} & AT+ENV & \yes & \no & \no & \yes & \yes & \yes & \partly \\
FullStack-Agent~\citep{fullstackagent} & AT & \yes & \no & \no & \yes & \no & \no & \no \\
\midrule
\textbf{\ourwork{} (ours)} & AT+ENV & \yes & \yes & \yes & \yes & \yes & \yes & \yes \\
\bottomrule
\end{tabular}}
\caption{Training resources for coding agents. Pair: code-supervision pairs; AT: agent trajectories; ENV: executable task environments. Runtime checks assess training solutions; public environments enable new attempts; records contain per-sample verification outcomes. \yes: documented coverage or release; \partly: partial coverage or release; \no: not established in the reviewed sources.}
\label{tab:related}
\end{table}

Building a web application requires a coding agent to implement interactive behavior and produce the intended interface. 
An application may resemble its target design while failing to respond correctly to user actions; conversely, a functional implementation may differ substantially from a visual reference. 
Assessing these requirements involves running the application, interacting with it in a browser, and inspecting its rendered appearance \citep{webgenbench,vision2web}. 
Front-end development therefore provides a setting in which agents must connect code changes with observable outcomes and revise their implementations when those outcomes fall short.

Training data for this process should capture how agents develop and check applications. Instruction--code and screenshot--code pairs provide supervision for producing implementations, but do not demonstrate the sequence of editing, execution, inspection and correction in a real completed application. Development trajectories expose these intermediate actions and the feedback available to the agent. Collecting such trajectories requires executable task environments: a development request, a designated starting workspace, applicable reference materials, and criteria for assessing the resulting application. Releasing these environments alongside trajectories also enables researchers to collect new attempts and verify their solutions.

Existing resources cover different parts of this need. 
Screenshot--code datasets provide visual supervision \citep{websight,web2code}, while software engineering and command-line pipelines construct executable tasks with tests \citep{swegym,cliuniverse}. 
Public resources for front-end development trajectories remain limited. 
WebGen-Agent releases a small collection generated from textual requests \citep{webgenagent}, while FullStack-Agent does not release its trajectories \citep{fullstackagent}. The training trajectories in both works address individual tasks specified in text, without covering development from visual references or iterative task sequences that build on earlier solutions. Table~\ref{tab:related} compares task coverage and released resources.

\begin{figure}[t]
\centering
\includegraphics[width=\textwidth]{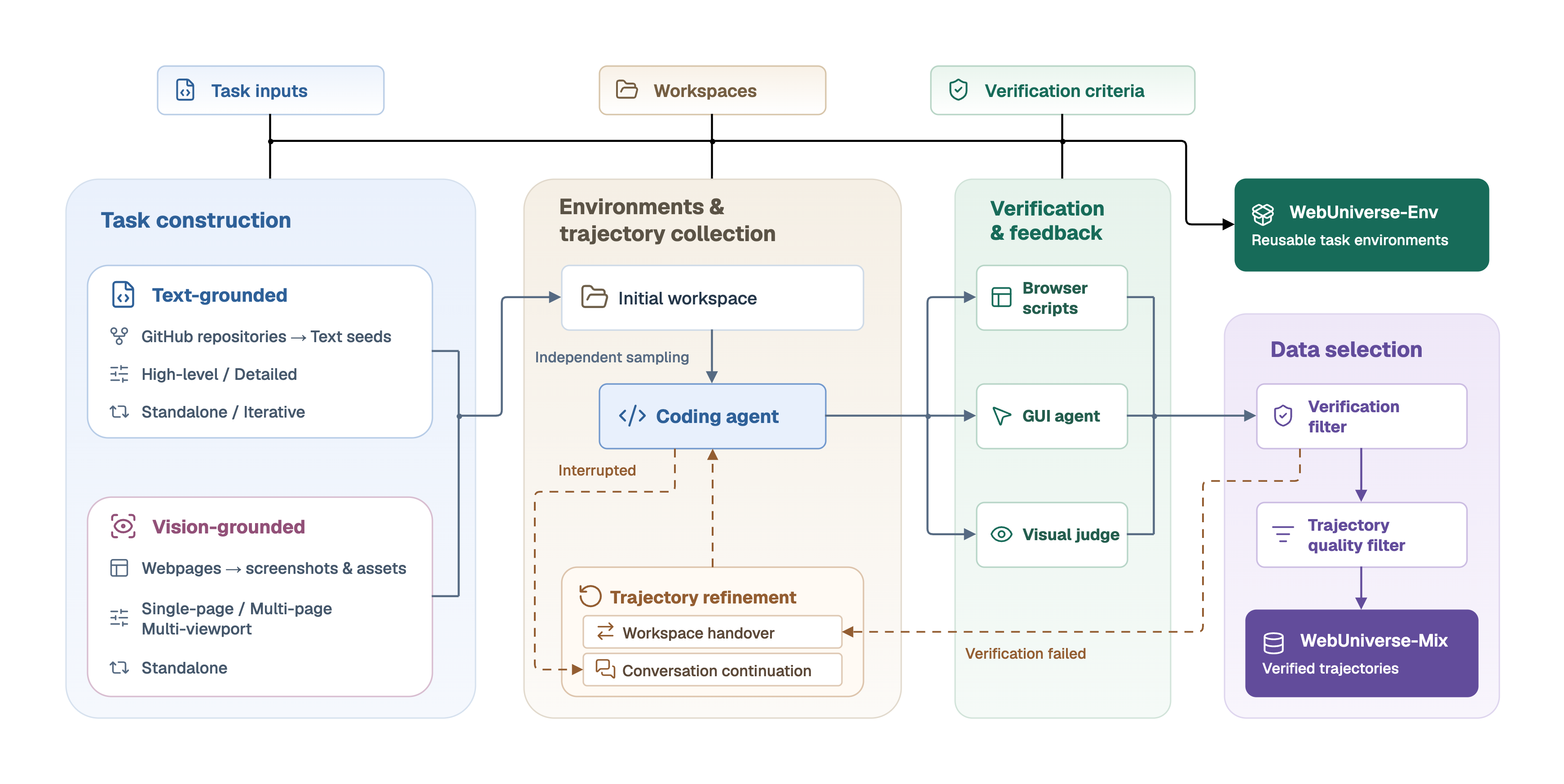}
\caption{The \ourwork{} construction pipeline. Independent sampling and trajectory refinement produce candidate trajectories, which undergo task-specific verification and quality filtering to form \ourdata{}. Task inputs, starting workspaces and verification criteria form \ourenv{}.}
\label{fig:pipeline}
\end{figure}

We introduce \ourwork{}, which provides 15K executable task environments for front-end web development (Figure~\ref{fig:pipeline}). We construct the task environments from open-source repositories and crawled pages. The \textbf{text-grounded} tasks ask an agent to build an application from written requirements, either as a standalone task or through a sequence of changes to the same codebase. The \textbf{vision-grounded} tasks ask the agent to reproduce real web pages from screenshots, including multiple pages and viewport widths.
Each environment preserves the task inputs, starting workspace and verification criteria, supporting further solving, verification and trajectory collection.

We collect trajectories through independent attempts and refinement of failed or interrupted attempts. 
Verification follows the requirements of each task: browser scripts or a GUI agent check application behavior, while a visual judge compares rendered pages with their references. We retain trajectories that pass task-specific verification and trajectory quality checks, and release each trajectory with its verification record and score. Together, the environments and trajectories provide both training demonstrations and the means to collect additional demonstrations under the same task requirements.

\ourmodel{} improves performance on Vision2Web \citep{vision2web}, WebGen-Bench \citep{webgenbench}, and Web-Bench \citep{webbench}, and outperforms StepFun-Mix \citep{stepfun} using the same starting checkpoint, training configuration and sample count.

Our contributions are: \textbf{(1) A scalable construction pipeline.} We introduce \ourwork{}, which constructs verifiable front-end development tasks from open-source repositories and real web pages, covering text-grounded development and vision-grounded reproduction. \textbf{(2) Reusable environments and verified trajectories.} We release executable task environments in \ourenv{}, together with 15K corresponding verified development trajectories in \ourdata{}, each accompanied by its verification record and score. \textbf{(3) Empirical validation and analysis.} We show improvements across three web-development benchmarks and over public-data training at the same sample budget. Our model approaches or exceeds frontier models on selected visual and functional metrics. Ablations demonstrate complementary benefits from text and visual trajectories and continued gains with increasing data size.

\section{Related Work}

\paragraph{Synthetic training data.} Large-scale synthetic datasets cover instruction following, reasoning, code generation, and tool use \citep{magpie,magicoder,opencodeinstruct,midtool,visualsphinx}. Some methods construct verifiable supervision through generated code tests \citep{kodcode,selfcodealign}. Software engineering corpora collect test-passing trajectories in executable repositories, for code issue resolution \citep{swesmith,swegym,r2egym}. Web-agent corpora instead collect interactions with existing sites \citep{agenttrek,osgenesis,nnetnav,explorer}. \ourwork{} targets the construction of web applications and checks the interfaces in a browser.

\paragraph{Task and environment synthesis.} ScaleEnv \citep{scaleenv} synthesizes interactive tool environments and tasks from scratch, validating environment code and executable action sequences. Agent World Model \citep{awm} synthesizes code-driven, database-backed tool environments for multi-turn agent reinforcement learning. CLI-Universe \citep{cliuniverse} builds containerized tasks and collects verified trajectories using tests and reference solutions. CodeMidas \citep{codemidas} turns implemented repository functionality into development tasks, grounded with tests. SWE-Flow \citep{sweflow} derives a development schedule from test dependencies and constructs a codebase and reference patch for each step.

\paragraph{Web page and web application generation.} WebSight and Web2Code pair screenshots with HTML in a single turn \citep{websight,web2code}. WebGen-Bench and WebGen-Agent report 600 and approximately 700 SFT trajectories, respectively, for text-specified website development \citep{webgenbench,webgenagent}. FullStack-Agent uses repository back-translation to synthesize development trajectories for full-stack applications \citep{fullstackagent}. The training examples in WebGen-Bench, WebGen-Agent and FullStack-Agent start from textual instructions and fresh workspaces. \ourwork{} releases 15K verified trajectories covering visual references and iterative development tasks, with per-sample verification records. Table~\ref{tab:related} compares task coverage, forms, verification and public releases.

\section{\ourwork{} Construction}
\label{sec:construction}
\label{sec:construction-overview}

A task specifies a development request to be completed from a designated starting workspace; a solving attempt is one agent execution on that task. \ourwork{} covers two task families (Figure~\ref{fig:pipeline}): text-grounded tasks define application behavior through written requirements, while vision-grounded tasks define the target appearance through reference screenshots. To increase task diversity, we construct multiple configurations within each family. Text-grounded tasks combine high-level or detailed specifications with standalone or iterative development; vision-grounded tasks cover single-page, multi-page and multi-viewport reproduction. We construct task statements and reference materials from open-source repositories and real websites (Section~\ref{sec:tasks}), prepare workspaces and collect solving trajectories (Section~\ref{sec:environments}), verify the resulting applications (Section~\ref{sec:verification}), and select passing trajectories for \ourdata{} (Section~\ref{sec:admission}).

\subsection{Task Construction}
\label{sec:tasks}
\label{sec:text}
\label{sec:seeds}

\paragraph{Text seeds.} We collect high-star repositories for front-end web applications through GitHub keyword searches and provide them to an agent. The agent reads the code and documentation, then summarizes the application's functionality, user actions and resulting interface changes. These application descriptions serve as \emph{text seeds} for task generation. For example, a jKanban seed describes adding cards and dragging them across board columns. We retain each seed's source repository name, URL and license.

\paragraph{Text-grounded tasks.} We use an LLM to generate high-level and detailed task statements from text seeds, changing the application scenario while preserving its underlying interaction mechanism. \emph{High-level specifications} state user goals, actions and expected interface feedback, leaving technical and design choices to the agent. \emph{Detailed specifications} additionally prescribe implementation and design details as needed, such as the technical stack, initial state, interface identifiers and appearance. We supply the selected scaffold's stack description when generating detailed tasks; for high-level tasks, we assume an empty workspace. Both levels specify concrete behavior implemented in the front end, without external APIs or backend services. At either level, we generate a \emph{standalone} development request or an \emph{iterative} sequence of requests. Standalone tasks ask for a complete application in one request. Iterative sequences model development under evolving requirements: the first task establishes the application, and later tasks extend or revise it. We generate all requests in a sequence jointly and explicitly state changes to earlier requirements.

\paragraph{Visual references.} We obtain candidate website URLs from FineWeb~\citep{fineweb}, screen their metadata, and check accessibility and crawling permissions. We visit eligible sites with a browser to collect page screenshots, supporting images and fonts, and source URLs. We apply programmatic checks to remove unusable screenshots and materials missing required images, followed by human review to select pages and sites. For multi-viewport references, we require usable desktop, tablet and mobile screenshots. During crawling, we also record navigation within each website: which link or button on one page opens another target page. Multi-page tasks use these records to specify the required navigation.

\paragraph{Vision-grounded tasks.} We assemble the selected materials into tasks using fixed templates, without calling a model. \emph{Single-page} tasks provide a reference screenshot and ask the agent to reproduce that page. \emph{Multi-page} tasks provide screenshots and navigation information for multiple target pages, asking the agent to reproduce both the pages and their navigation. \emph{Multi-viewport} tasks provide reference screenshots of the same page at desktop, tablet and mobile widths and require matching its appearance at each width. We use the standalone setting for vision-grounded tasks, as collecting visual references for intermediate development states is challenging for websites in the wild. We defer further details on source processing, task specifications and scaffold configurations to Appendix~\ref{app:supplementary}.

\subsection{Environments and Trajectory Collection}
\label{sec:environments}
\label{sec:modes}
\label{sec:env}

\paragraph{Initial workspaces.} We provide a shared container runtime with development tools and a browser for all task types. Detailed tasks start from the project scaffold supplied during task generation; high-level tasks start from an empty application workspace. For vision-grounded tasks, we provide reference screenshots and supporting assets, together with navigation information for multi-page tasks. We use Harbor to manage containers and workspaces and OpenCode~\citep{opencode} as the agent harness. Our project scaffolds cover different front-end frameworks, state-management libraries, styling tools and bundlers.

\paragraph{Trajectory collection.} To collect diverse demonstrations of task completion, error correction and recovery, we use a \emph{trajectory model pool} with two collection methods. \textbf{(1) Independent sampling.} We start each attempt in a fresh conversation from the task's designated initial workspace, without importing dialogue or feedback from another attempt. The agent receives the task statement, edits code, executes commands and inspects the application before submitting it for verification. This method applies to standalone tasks and individual tasks within iterative sequences. After a task passes verification, we save its project files as the next task's starting workspace, allowing subsequent tasks to be sampled independently. We stop a sequence when no attempt passes the current task. \textbf{(2) Trajectory refinement.} Successful independent attempts may contain few demonstrations of correcting errors or recovering from stalled execution. We collect such behavior by continuing from failed or interrupted attempts with the same model or a different model, using verifier feedback when available. \emph{Workspace handover} starts a new conversation with the task statement, saved failed workspace and any available verifier feedback. \emph{Conversation continuation} preserves an actively interrupted attempt's workspace and prior dialogue, together with any feedback already received. We reuse the same saved starting state for repeated handover attempts. Continuation samples retain the initial agent's history as context; we exclude its assistant content and tool calls from the training loss and supervise only the resumed portion. Appendices~\ref{app:scaffold} and~\ref{app:batch} detail the agent harness and continuation settings.

\begin{figure}[t]
\centering
\includegraphics[width=0.45\linewidth]{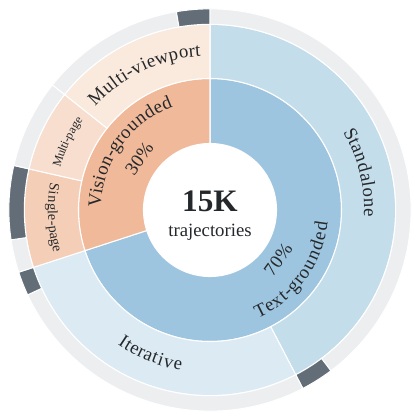}
\caption{Composition of \ourdata{} (15K trajectories). Rings encode task family, variant and collection mode, from inside to outside. The outer ring distinguishes independent sampling (light gray) from trajectory refinement (dark gray). Iterative counts represent trajectories for individual tasks.}
\label{fig:composition}
\end{figure}

\subsection{Verification and Feedback}
\label{sec:verification}

We select a verifier according to the task requirements to check the application's functionality or appearance.

\paragraph{Browser scripts.} We verify detailed tasks with browser scripts. Before solving, we use an LLM to extract verification criteria and passing conditions from the task statement and generate test code without access to a solution. The scripts check explicit source-code requirements and use Playwright to verify specified interactions and appearance. Generated tests may impose assumptions not specified in the task statement, causing false negatives. We therefore use an LLM to review only failed tests for such assumptions. We regenerate faulty tests and rerun them on the unchanged application.

\paragraph{GUI agent.} We use a GUI agent to verify high-level tasks and some detailed tasks, adapting to different implementations. The agent explores the running application and executes tests through the browser based on the task statement, without reading source code. It records \emph{pass}, \emph{fail} or \emph{blocked} with supporting observations for each test.

\paragraph{Visual judge.} We use a VLM as a visual judge to compare each target page's rendered screenshot with its reference at the specified viewport, using a six-level rubric. For multi-page tasks, the verifier opens target pages using their specified routes or navigation paths; unreachable pages score zero.

\paragraph{Verification records and feedback.} We apply the task-specific verifier to the application produced by each collected trajectory. We retain outcomes, scores and supporting evidence for data selection and use available failure feedback for trajectory refinement.

\subsection{Data Selection}
\label{sec:admission}

We select trajectories through two filters. \textbf{(1) Verification filter.} Verification must complete normally, produce all required results and satisfy the task-specific passing conditions. Browser-script and GUI-agent verification require all tests to pass; browser-script review must also finish without unresolved issues. Visual verification requires every page--viewport comparison to reach at least \emph{good match} ($0.75$). \textbf{(2) Trajectory quality filter.} We exclude samples if conversation or image records are incomplete or the model does not end the final solving session normally. We check record structure, image references and compatibility with the training format. For iterative task sequences, we retain a consecutive passing prefix beginning with the first task and containing at least three tasks, reducing overrepresentation of early tasks.

\paragraph{Decontamination.}
\label{sec:decontam}
Before release, we screen selected tasks for benchmark overlap using task-statement similarity for text-grounded tasks and source websites for vision-grounded tasks. We exclude confirmed overlaps and retain clear false positives. Appendix~\ref{app:decontamination} details the screening rules, review outcomes and subset statistics.

\begin{table}[!t]
\centering
\footnotesize
\setlength{\tabcolsep}{.4pt}
\renewcommand{\arraystretch}{1.15}
\begin{tabularx}{\textwidth}{@{}l*{11}{>{\centering\arraybackslash}X}@{}}
\toprule
\multirow{2}{*}{Model} & \multicolumn{4}{c}{Vision2Web} & \multicolumn{4}{c}{WebGen-Bench} & \multicolumn{3}{c}{Web-Bench} \\
\cmidrule(lr){2-5}\cmidrule(lr){6-9}\cmidrule(l){10-12}
 & Desktop & Tablet & Mobile & Avg. & Func. & Data & Design & Overall & {\scriptsize PA@1} & {\scriptsize PA@2} & Avg. \\
\midrule
\modelicon{\includegraphics[width=.9em,height=.9em,keepaspectratio]{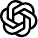}} GPT-5.6 Sol & 77.2 & 77.1 & 73.0 & 75.8 & 51.6 & \underline{72.6} & 87.7 & 64.5 & 23.5 & 62.9 & 43.2 \\
\modelicon{\includegraphics[width=.9em,height=.9em,keepaspectratio]{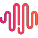}} MiniMax-M3 & 74.0 & \underline{67.4} & \underline{61.9} & \underline{67.8} & \underline{48.1} & 76.3 & 86.1 & \underline{63.4} & 25.1 & 49.4 & 37.3 \\
\modelicon{\includegraphics[width=.9em,height=.9em,keepaspectratio]{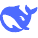}} DPSK-V4-Flash-Vision-Exp & 77.2 & 74.6 & 71.4 & 74.4 & \underline{49.0} & \underline{68.8} & \underline{82.0} & \underline{60.9} & 22.0 & 39.1 & 30.6 \\
\modelicon{\includegraphics[width=.9em,height=.9em,keepaspectratio]{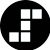}} Step-3.7-Flash & 74.4 & 69.7 & 65.3 & 69.8 & \underline{50.7} & \underline{71.5} & \underline{80.3} & \underline{62.3} & 23.1 & 43.3 & 33.2 \\
\modelicon{\includegraphics[width=.9em,height=.9em,keepaspectratio]{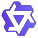}} Qwen3.5-27B & \underline{63.9} & \underline{57.0} & \underline{52.1} & \underline{57.7} & \underline{25.1} & \underline{53.0} & \underline{70.1} & \underline{41.6} & \underline{13.3} & 23.3 & \underline{18.3} \\
\midrule
\modelicon{\includegraphics[width=.9em,height=.9em,keepaspectratio]{figures/model-icons/qwen-color.pdf}} Qwen3.5-35B-A3B & 57.1 & 51.0 & 46.8 & 51.7 & 22.3 & 44.1 & 53.7 & 34.5 & 7.2 & 12.2 & 9.7 \\
\quad + StepFun-Mix & 58.0 & 51.9 & 47.3 & 52.4 & 21.2 & 39.0 & 48.8 & 31.5 & 9.3 & 17.7 & 13.5 \\
\rowcolor[RGB]{235,250,235}
\quad \textbf{+ \ourdata{}} & 72.6 & \textbf{69.4} & \textbf{62.9} & \textbf{68.3} & \textbf{51.3} & \textbf{73.7} & \textbf{84.0} & \textbf{63.9} & 13.9 & 23.1 & 18.5 \\
\bottomrule
\end{tabularx}
\caption{Main benchmark results; Vision2Web reports Level~1 scores. The + rows fine-tune Qwen3.5-35B-A3B on 15K samples from StepFun-Mix or \ourdata{}. Func., Data and Design denote functional, data-display and design-validation accuracy; DPSK denotes DeepSeek and PA denotes pass. WebGen-Bench and Web-Bench report percentages; Web-Bench Avg. averages PA@1 and PA@2. Higher is better. Bold \ourdata{} scores exceed at least one of the first four reference models; underlined reference scores are below ours.}
\label{tab:main}
\end{table}

\subsection{Dataset Analysis}
\label{sec:composition}

\paragraph{Artifacts.} Our final artifacts comprise \ourdata{}, a dataset of 15K verified development trajectories, and \ourenv{}, a set of 15K corresponding reusable task environments. Each task environment preserves the starting workspace, reference materials and verification criteria, supporting trajectory rollout and collection. \ourdata{} comprises 10,500 text-grounded and 4,500 vision-grounded trajectories (Figure~\ref{fig:composition}). Task-package details are documented in Appendix~\ref{app:env}.

\paragraph{Coverage and trajectory characteristics.} Trajectories in \ourdata{} cover 1,559 open-source repositories and 3,500 website sources. Initial workspaces range from empty applications to framework-specific scaffolds. React is the most common framework observed in text-grounded trajectories (65.3\%), while plain HTML/CSS/JS predominates in vision-grounded trajectories (87.8\%). The trajectories contain 511.9M tokens, 551,818 tool calls and 74,811 image occurrences. Text-grounded trajectories have medians of 20.6K tokens and 20 tool calls, compared with 59.4K tokens and 70 tool calls for vision-grounded trajectories. We report additional coverage and trajectory statistics in Appendix~\ref{app:statistics}.

\paragraph{Quality audit.}
\label{sec:audit}
We conduct a human audit of 100 accepted and 100 rejected attempts per verifier from the collection pool (600 in total), examining task statements, delivered applications and verification records. We report accuracy, precision, recall and F1 for all three verifiers. Appendix~\ref{app:audit} provides the audit protocol and results.

\section{Experiments}
\label{sec:experiments}

\begin{table}[t]
\centering
\footnotesize
\setlength{\tabcolsep}{1pt}
\renewcommand{\arraystretch}{1.15}
\begin{tabularx}{\textwidth}{@{}l*{11}{>{\centering\arraybackslash}X}@{}}
\toprule
\multirow{2}{*}{Model} & \multicolumn{4}{c}{Vision2Web} & \multicolumn{4}{c}{WebGen-Bench} & \multicolumn{3}{c}{Web-Bench} \\
\cmidrule(lr){2-5}\cmidrule(lr){6-9}\cmidrule(l){10-12}
 & Desktop & Tablet & Mobile & Avg. & Func. & Data & Design & Overall & {\scriptsize PA@1} & {\scriptsize PA@2} & Avg. \\
\midrule
\modelicon{\includegraphics[width=.9em,height=.9em,keepaspectratio]{figures/model-icons/qwen-color.pdf}} Qwen3.5-35B-A3B & 57.1 & 51.0 & 46.8 & 51.7 & 22.3 & 44.1 & 53.7 & 34.5 & 7.2 & 12.2 & 9.7 \\
\quad + Text-grounded subset & 58.2 & 54.7 & 49.2 & 54.1 & \underline{42.8} & \underline{65.1} & \underline{79.5} & \underline{56.1} & \underline{13.5} & \underline{18.1} & \underline{15.8} \\
\quad + Vision-grounded subset & \underline{66.6} & \underline{65.6} & \underline{59.4} & \underline{63.9} & 6.0 & 15.9 & 23.4 & 12.1 & 5.3 & 6.8 & 6.1 \\
\quad \textbf{+ \ourdata{}} & \textbf{72.6} & \textbf{69.4} & \textbf{62.9} & \textbf{68.3} & \textbf{51.3} & \textbf{73.7} & \textbf{84.0} & \textbf{63.9} & \textbf{13.9} & \textbf{23.1} & \textbf{18.5} \\
\bottomrule
\end{tabularx}
\caption{Data composition ablation of \ourdata{} using the same checkpoint and training configuration. Metrics follow Table~\ref{tab:main}; bold and underlined values indicate the best and second-best scores, respectively.}
\label{tab:data-ablation}
\end{table}

\begin{figure}[t]
\centering
\captionsetup[subfigure]{font=small,skip=3pt,justification=centering}
\begin{subfigure}[b]{0.32\textwidth}
\centering
\includegraphics[width=\linewidth]{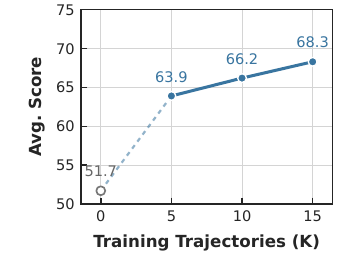}
\caption{Vision2Web}
\label{fig:scaling-v2w}
\end{subfigure}\hfill
\begin{subfigure}[b]{0.32\textwidth}
\centering
\includegraphics[width=\linewidth]{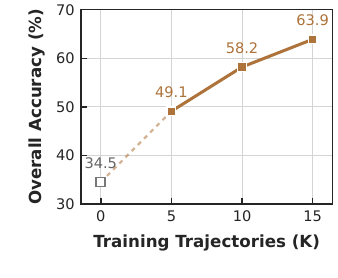}
\caption{WebGen-Bench}
\label{fig:scaling-wgb}
\end{subfigure}\hfill
\begin{subfigure}[b]{0.32\textwidth}
\centering
\includegraphics[width=\linewidth]{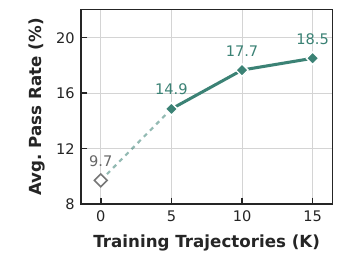}
\caption{Web-Bench}
\label{fig:scaling-webbench}
\end{subfigure}
\caption{Results by training-data size. Open markers at 0 denote Qwen3.5-35B-A3B; 15K denotes \ourmodel{}. Panels use different y-axis ranges with nonzero origins. Higher is better.}
\label{fig:scaling}
\end{figure}

\subsection{Setup}
\label{sec:setup}

\paragraph{Training.} We fine-tune Qwen3.5-35B-A3B on the 15K trajectories in \ourdata{} to obtain \ourmodel{}. We update all parameters except the vision tower over two epochs and apply loss to assistant outputs. All fine-tuning runs use the same hyperparameters; Appendix~\ref{app:training-hyperparameters} provides the full configuration.

\paragraph{Baselines.} We compare \ourmodel{} with the original Qwen3.5-35B-A3B to measure gains from fine-tuning. For a public-data comparison at a matched 15K sample budget, we fine-tune the same checkpoint on frontier-level web development oriented \textit{StepFun-Mix}, a 15K subset of the public \textit{Step-3.5-Flash-SFT} dataset~\citep{stepfun}, using the same training configuration. We call this baseline Qwen3.5-StepFun; Appendix~\ref{app:stepfun-data} describes the subset construction. Table~\ref{tab:main} also reports results for additional reference models.

\paragraph{Benchmarks.} We evaluate page reproduction, functional development and iterative development. Vision2Web (Level~1)~\citep{vision2web} measures visual reproduction on 100 tasks at desktop, tablet and mobile widths. WebGen-Bench~\citep{webgenbench} evaluates 101 applications built from text requirements through functional, data-display and design-validation checks. Web-Bench~\citep{webbench} evaluates 50 projects with 20 successive development tasks each; we report pass@1, pass@2 and their arithmetic mean. Supplementary results on Vision2Web (Level~2) are reported in Appendix~\ref{app:v2w-l2}.

\paragraph{Evaluation protocol.} We run all models in OpenCode with their recommended inference settings. We use GPT-5 for Vision2Web visual grading, GPT-5.4 for WebGen-Bench, and the official test scripts for Web-Bench, keeping each benchmark's evaluator fixed across models. We report one run per model and benchmark in the main experiments. Appendix~\ref{app:eval-setup} details the evaluation protocol and score aggregation.

\begin{figure}[t]
\centering
\includegraphics[width=0.95\textwidth]{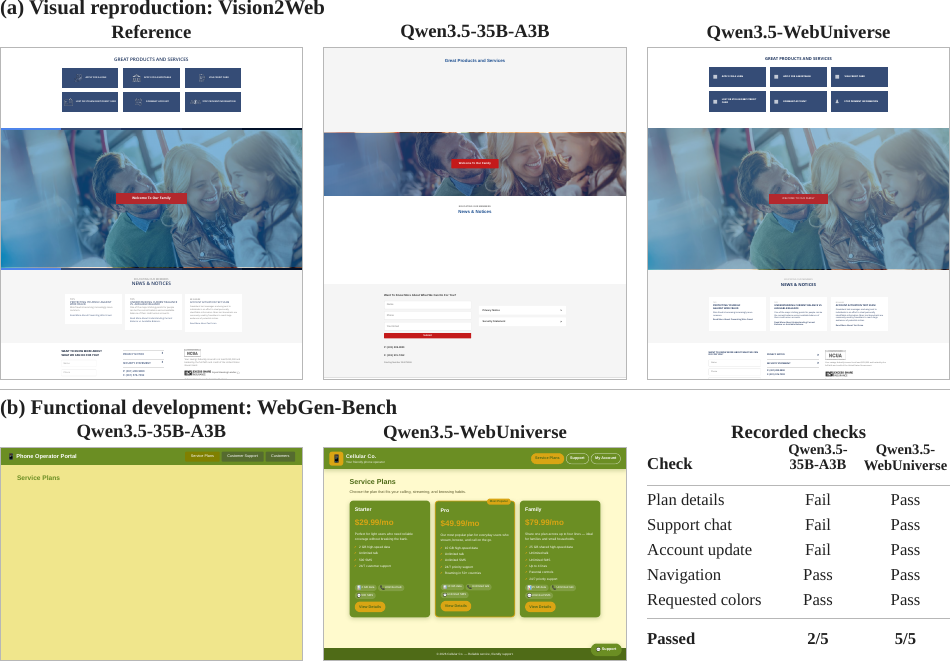}
\caption{Qualitative examples: (a) corresponding lower-page sections on Vision2Web's tefcu task; (b) phone-operator applications and recorded checks on WebGen-Bench task 000020.}
\label{fig:qualitative-cases}
\end{figure}

\subsection{Main Results}

\paragraph{\ourdata{} improves agentic front-end web development.} Fine-tuning improves visual reproduction, functional development and iterative development (Table~\ref{tab:main}). The gains span all viewport sizes and functional check categories, covering both appearance and interactive behavior. Higher Web-Bench pass rates also indicate better performance on successive requests within an existing application. Our 35B model also approaches or exceeds larger frontier models on selected visual and functional metrics. For example, it scores 63.9\% on WebGen-Bench versus 62.3\% for Step-3.7-Flash (198B total parameters; \citealp{step37flash}), with Vision2Web averages of 68.3 and 69.8, respectively.

\paragraph{\ourdata{} outperforms matched public-data training.} With the same starting checkpoint, training configuration and sample count, \ourdata{} exceeds StepFun-Mix on every reported metric. The advantage is particularly pronounced in functional development, where the public mixture underperforms the original model. This contrast supports collecting trajectories aligned with the target development tasks.

\paragraph{Fine-tuning is accompanied by more frequent checks.} Supplementary traces show more frequent checks after the final edit, through local execution or file rereading. Functional-development traces also contain more error--edit--recheck cycles. Repetition and retries without implementation changes nevertheless persist. Appendix~\ref{app:behavior} provides the definitions, statistics and representative cases supporting these workflow observations.

\paragraph{Improvements extend to page structure and user interactions.} Figure~\ref{fig:qualitative-cases} illustrates the gains in delivered applications. The fine-tuned visual output more closely follows the reference's section layout and assets. The phone-operator application passes checks for plan details, support chat and account updates that fail in the original model's output. Appendix~\ref{app:qualitative} shows the complete pages for the visual example.

\subsection{Ablation}
\label{sec:ablation}
\label{sec:data-ablation}
\label{sec:scaling}

\paragraph{Text and visual trajectories support complementary capabilities.} We fine-tune the same checkpoint on each task family separately and on their full mixture, using the shared training configuration (Table~\ref{tab:data-ablation}). Text-grounded training primarily benefits functional and iterative development, with limited gains in visual reproduction. Vision-grounded training substantially improves visual reproduction, while functional accuracy and iterative pass rates fall below the original model. The mixture achieves the highest scores on every reported metric, supporting coverage of both appearance and behavior.

\paragraph{Performance continues to improve as the dataset grows.} We fine-tune on strictly nested 5K, 10K and 15K subsets of \ourdata{}, fixing the mixture proportions and training configuration. Figure~\ref{fig:scaling} shows monotonic improvements across all three benchmarks. Visual reproduction benefits especially from the first subset, while functional accuracy continues to rise substantially with more data. Additional trajectories are still valuable at the largest evaluated size, even with diminishing incremental gains.

\section{Conclusion}

We introduced \ourwork{}, a pipeline for scaling verified training data for front-end web development from open-source repositories and real web pages.
The resulting release pairs 15K verified trajectories in \ourdata{} with their corresponding executable task environments in \ourenv{}, covering text-grounded development and vision-grounded reproduction.
Fine-tuning improves visual reproduction, functional and iterative development, outperforming matched public-data training across all reported metrics.
Ablations show complementary benefits from text and visual trajectories and continued improvements with increasing data size.
Together, the trajectories, verification records and reusable environments support agent training and further trajectory rollout and collection.

\section*{Limitations}

\paragraph{Task domains.} Our current task collection focuses on web front-end development. Extending task construction and verification to native mobile interfaces, 3D applications and games is a direction for future work.

\paragraph{Visual references.} Vision-grounded tasks specify target appearance through screenshots at selected viewports. These references do not explicitly capture temporal visual behavior, such as animations and transitions. Future work could extend visual references with videos or interaction recordings.

\bibliographystyle{iclr2027_conference}
\bibliography{refs}

@inproceedings{kodcode,
  title={{KodCode}: A Diverse, Challenging, and Verifiable Synthetic Dataset for Coding},
  author={Xu, Zhangchen and Liu, Yang and Yin, Yueqin and Zhou, Mingyuan and Poovendran, Radha},
  booktitle={Findings of the {A}ssociation for {C}omputational {L}inguistics: {ACL} 2025},
  pages={6980--7008},
  year={2025},
  publisher = {Association for Computational Linguistics},
  doi       = {10.18653/v1/2025.findings-acl.365},
  url       = {https://aclanthology.org/2025.findings-acl.365/}
}

@inproceedings{magicoder,
  title     = {{Magicoder}: Empowering Code Generation with {OSS-Instruct}},
  author    = {Wei, Yuxiang and Wang, Zhe and Liu, Jiawei and Ding, Yifeng and Zhang, Lingming},
  booktitle = {Proceedings of the 41st International Conference on Machine Learning},
  year      = {2024},
  pages     = {52632--52657},
  volume    = {235},
  series    = {Proceedings of Machine Learning Research},
  publisher = {PMLR},
  url       = {https://proceedings.mlr.press/v235/wei24h.html}
}

@inproceedings{selfcodealign,
  title     = {{SelfCodeAlign}: Self-Alignment for Code Generation},
  author    = {Wei, Yuxiang and Cassano, Federico and Liu, Jiawei and Ding, Yifeng and Jain, Naman and Mueller, Zachary and de Vries, Harm and von Werra, Leandro and Guha, Arjun and Zhang, Lingming},
  booktitle = {Advances in Neural Information Processing Systems},
  year      = {2024},
  pages     = {62787--62874},
  volume    = {37},
  publisher = {Curran Associates, Inc.},
  doi       = {10.52202/079017-2008},
  url       = {https://proceedings.neurips.cc/paper_files/paper/2024/hash/72da102da91a8042a0b2aa968429a9f9-Abstract-Conference.html}
}

@misc{opencodeinstruct,
  title     = {{OpenCodeInstruct}: A Large-scale Instruction Tuning Dataset for Code {LLMs}},
  author    = {Ahmad, Wasi Uddin and Ficek, Aleksander and Samadi, Mehrzad and Huang, Jocelyn and Noroozi, Vahid and Majumdar, Somshubra and Ginsburg, Boris},
  year      = {2025},
  eprint    = {2504.04030},
  archivePrefix = {arXiv},
  url       = {https://arxiv.org/abs/2504.04030}
}

@inproceedings{swesmith,
  title     = {{SWE-smith}: Scaling Data for Software Engineering Agents},
  author    = {Yang, John and Lieret, Kilian and Jimenez, Carlos E. and Wettig, Alexander and Khandpur, Kabir and Zhang, Yanzhe and Hui, Binyuan and Press, Ofir and Schmidt, Ludwig and Yang, Diyi},
  booktitle = {Advances in Neural Information Processing Systems},
  year      = {2025},
  volume    = {38},
  publisher = {Curran Associates, Inc.},
  doi       = {10.52202/085713-3239},
  url       = {https://proceedings.neurips.cc/paper_files/paper/2025/hash/8b86cf5ace600c48fd188efbb8dedec8-Abstract-Datasets_and_Benchmarks_Track.html}
}

@inproceedings{swegym,
  title     = {Training Software Engineering Agents and Verifiers with {SWE-Gym}},
  author    = {Pan, Jiayi and Wang, Xingyao and Neubig, Graham and Jaitly, Navdeep and Ji, Heng and Suhr, Alane and Zhang, Yizhe},
  booktitle = {Proceedings of the 42nd International Conference on Machine Learning},
  year      = {2025},
  pages     = {47717--47737},
  volume    = {267},
  series    = {Proceedings of Machine Learning Research},
  publisher = {PMLR},
  url       = {https://proceedings.mlr.press/v267/pan25g.html}
}

@inproceedings{r2egym,
  title     = {{R2E-Gym}: Procedural Environment Generation and Hybrid Verifiers for Scaling Open-Weights {SWE} Agents},
  author    = {Jain, Naman and Singh, Jaskirat and Shetty, Manish and Zhang, Tianjun and Zheng, Liang and Sen, Koushik and Stoica, Ion},
  booktitle = {Second Conference on Language Modeling},
  year      = {2025},
  url       = {https://openreview.net/forum?id=7evvwwdo3z}
}

@inproceedings{sweflow,
  title     = {{SWE-Flow}: Synthesizing Software Engineering Data in a Test-Driven Manner},
  author    = {Zhang, Lei and Yang, Jiaxi and Yang, Min and Yang, Jian and Chen, Mouxiang and Zhang, Jiajun and Cui, Zeyu and Hui, Binyuan and Lin, Junyang},
  booktitle = {Proceedings of the 42nd International Conference on Machine Learning},
  pages     = {76518--76540},
  volume    = {267},
  series    = {Proceedings of Machine Learning Research},
  year      = {2025},
  url       = {https://proceedings.mlr.press/v267/zhang25cn.html},
  publisher = {PMLR}
}

@inproceedings{agenttrek,
  title     = {{AgentTrek}: Agent Trajectory Synthesis via Guiding Replay with Web Tutorials},
  author    = {Xu, Yiheng and Lu, Dunjie and Shen, Zhennan and Wang, Junli and Wang, Zekun and Mao, Yuchen and Xiong, Caiming and Yu, Tao},
  booktitle = {International Conference on Learning Representations},
  year      = {2025},
  pages     = {79822--79843},
  url       = {https://proceedings.iclr.cc/paper_files/paper/2025/hash/c681fb2bf1d785fbc766f3ea14758aab-Abstract-Conference.html}
}

@misc{nnetnav,
  title     = {{NNetNav}: Unsupervised Learning of Browser Agents Through Environment Interaction in the Wild},
  author    = {Murty, Shikhar and Zhu, Hao and Bahdanau, Dzmitry and Manning, Christopher D.},
  year      = {2024},
  eprint    = {2410.02907},
  archivePrefix = {arXiv},
  url       = {https://arxiv.org/abs/2410.02907}
}

@inproceedings{explorer,
  title     = {{Explorer}: Scaling Exploration-driven Web Trajectory Synthesis for Multimodal Web Agents},
  author    = {Pahuja, Vardaan and Lu, Yadong and Rosset, Corby and Gou, Boyu and Mitra, Arindam and Whitehead, Spencer and Su, Yu and Awadallah, Ahmed Hassan},
  booktitle = {Findings of the {A}ssociation for {C}omputational {L}inguistics: {ACL} 2025},
  year      = {2025},
  pages     = {6300--6323},
  publisher = {Association for Computational Linguistics},
  doi       = {10.18653/v1/2025.findings-acl.326},
  url       = {https://aclanthology.org/2025.findings-acl.326/}
}

@misc{websight,
  title     = {Unlocking the Conversion of Web Screenshots into {HTML} Code with the {WebSight} Dataset},
  author    = {Lauren{\c{c}}on, Hugo and Tronchon, L{\'e}o and Sanh, Victor},
  year      = {2024},
  eprint    = {2403.09029},
  archivePrefix = {arXiv},
  url       = {https://arxiv.org/abs/2403.09029}
}

@inproceedings{web2code,
  title     = {{Web2Code}: A Large-scale Webpage-to-Code Dataset and Evaluation Framework for Multimodal {LLMs}},
  author    = {Yun, Sukmin and Lin, Haokun and Thushara, Rusiru and Bhat, Mohammad Qazim and Wang, Yongxin and Jiang, Zutao and Deng, Mingkai and Wang, Jinhong and Tao, Tianhua and Li, Junbo and Li, Haonan and Nakov, Preslav and Baldwin, Timothy and Liu, Zhengzhong and Xing, Eric P. and Liang, Xiaodan and Shen, Zhiqiang},
  booktitle = {Advances in Neural Information Processing Systems},
  year      = {2024},
  pages     = {112134--112157},
  volume    = {37},
  publisher = {Curran Associates, Inc.},
  doi       = {10.52202/079017-3560},
  url       = {https://proceedings.neurips.cc/paper_files/paper/2024/hash/cb66be286795d71f89367d596bf78ea7-Abstract.html}
}

@inproceedings{vision2web,
  title     = {{Vision2Web}: A Hierarchical Benchmark for Visual Website Development with Agent Verification},
  author    = {He, Zehai and Hong, Wenyi and Yang, Zhen and Pan, Ziyang and Liu, Mingdao and Gu, Xiaotao and Tang, Jie},
  booktitle = {Proceedings of the 43rd International Conference on Machine Learning},
  year      = {2026},
  pages     = {41393--41415},
  volume    = {306},
  series    = {Proceedings of Machine Learning Research},
  publisher = {PMLR},
  url       = {https://proceedings.mlr.press/v306/he26n.html}
}

@inproceedings{webgenbench,
  title     = {{WebGen-Bench}: Evaluating {LLMs} on Generating Interactive and Functional Websites from Scratch},
  author    = {Lu, Zimu and Yang, Yunqiao and Ren, Houxing and Hou, Haotian and Xiao, Han and Wang, Ke and Shi, Weikang and Zhou, Aojun and Zhan, Mingjie and Li, Hongsheng},
  booktitle = {Advances in Neural Information Processing Systems},
  year      = {2025},
  volume    = {38},
  publisher = {Curran Associates, Inc.},
  doi       = {10.52202/085713-2419},
  url       = {https://proceedings.neurips.cc/paper_files/paper/2025/hash/6841eed8bb6a2ec49e49235c8115efee-Abstract-Datasets_and_Benchmarks_Track.html}
}

@inproceedings{webgenagent,
  title     = {{WebGen-Agent}: Enhancing Interactive Website Generation with Multi-Level Feedback and Step-Level Reinforcement Learning},
  author    = {Lu, Zimu and Ren, Houxing and Yang, Yunqiao and Wang, Ke and Zong, Zhuofan and Pan, Junting and Zhan, Mingjie and Li, Hongsheng},
  booktitle = {International Conference on Learning Representations},
  year      = {2026},
  pages     = {83478--83525},
  url       = {https://proceedings.iclr.cc/paper_files/paper/2026/hash/86c408308bec72ecce4d86edf243634c-Abstract-Conference.html}
}

@inproceedings{fullstackagent,
  title     = {{FullStack-Agent}: Enhancing Agentic Full-Stack Web Coding via Development-Oriented Testing and Repository Back-Translation},
  author    = {Lu, Zimu and Ren, Houxing and Yang, Yunqiao and Wang, Ke and Zong, Zhuofan and Zhan, Mingjie and Li, Hongsheng},
  booktitle = {Proceedings of the 43rd International Conference on Machine Learning},
  year      = {2026},
  pages     = {81478--81512},
  volume    = {306},
  series    = {Proceedings of Machine Learning Research},
  publisher = {PMLR},
  url       = {https://proceedings.mlr.press/v306/lu26ab.html}
}

@misc{webbench,
  title     = {{Web-Bench}: A {LLM} Code Benchmark Based on Web Standards and Frameworks},
  author    = {Xu, Kai and Mao, YiWei and Guan, XinYi and Feng, ZiLong},
  year      = {2025},
  eprint    = {2505.07473},
  archivePrefix = {arXiv},
  url       = {https://arxiv.org/abs/2505.07473}
}

@inproceedings{fineweb,
  title     = {The {FineWeb} Datasets: Decanting the Web for the Finest Text Data at Scale},
  author    = {Penedo, Guilherme and Kydl{\'i}{\v{c}}ek, Hynek and Ben allal, Loubna and Lozhkov, Anton and Mitchell, Margaret and Raffel, Colin and Von Werra, Leandro and Wolf, Thomas},
  booktitle = {Advances in Neural Information Processing Systems},
  year      = {2024},
  pages     = {30811--30849},
  volume    = {37},
  publisher = {Curran Associates, Inc.},
  doi       = {10.52202/079017-0970},
  url       = {https://proceedings.neurips.cc/paper_files/paper/2024/hash/370df50ccfdf8bde18f8f9c2d9151bda-Abstract-Datasets_and_Benchmarks_Track.html}
}

@inproceedings{osgenesis,
  title     = {{OS-Genesis}: Automating {GUI} Agent Trajectory Construction via Reverse Task Synthesis},
  author    = {Sun, Qiushi and Cheng, Kanzhi and Ding, Zichen and Jin, Chuanyang and Wang, Yian and Xu, Fangzhi and Wu, Zhenyu and Chen, Liheng and Jia, Chengyou and Liu, Zhoumianze and Kao, Ben and Li, Guohao and He, Junxian and Qiao, Yu and Wu, Zhiyong},
  booktitle = {Proceedings of the 63rd Annual Meeting of the {A}ssociation for {C}omputational {L}inguistics (Volume 1: Long Papers)},
  year      = {2025},
  pages     = {5555--5579},
  publisher = {Association for Computational Linguistics},
  doi       = {10.18653/v1/2025.acl-long.277},
  url       = {https://aclanthology.org/2025.acl-long.277/}
}

@misc{opencode,
  title     = {{OpenCode}: an open-source coding agent},
  author    = {{OpenCode contributors}},
  year      = {2026},
  howpublished = {\url{https://github.com/anomalyco/opencode}},
  note      = {Version 1.18.11 (npm package \texttt{opencode-ai})}
}

@misc{stepfun,
  title     = {{Step-3.5-Flash-SFT}},
  author    = {{StepFun}},
  year      = {2026},
  howpublished = {\url{https://huggingface.co/datasets/stepfun-ai/Step-3.5-Flash-SFT}},
  note      = {Dataset release}
}

@misc{cliuniverse,
  title     = {{CLI-Universe}: Towards Verifiable Task Synthesis Engine for Terminal Agents},
  author    = {Hua, Zhanbo and Yao, Yifan and Xie, Weihao and Zhao, Yongchi and Liu, Minghao and Qiu, Ruizhi and Huang, Zhewei and Wang, Zun and Ji, Yiyan and Ye, Yunhai and Zhu, Letian and Lei, Xinping and Li, Han and Ma, Zhiyuan and Wang, Zili and Zhang, Zhaoxiang and Liu, Jiaheng},
  year      = {2026},
  eprint    = {2606.22883},
  archivePrefix = {arXiv},
  url       = {https://arxiv.org/abs/2606.22883}
}

@misc{codemidas,
  title     = {{CodeMidas}: Scaling Agentic Coding {RL} Environments from Code Itself},
  author    = {Ye, Bowen and Li, Lei and Li, Shicheng and Yue, Zihao and Zhang, Linghao and Lv, Hanglong and Liu, Yuanxin and Ma, Wenhan and Tian, Hao and Li, Rang and Dong, Jinhao and Zhao, Yikai and Deng, Xiangwei and Zhang, Hailin and Zhao, Liang and Liu, Qi and Kong, Lingpeng and Yang, Tong and Luo, Fuli},
  year      = {2026},
  eprint    = {2609.22068},
  archivePrefix = {arXiv},
  url       = {https://arxiv.org/abs/2609.22068}
}

@inproceedings{magpie,
  author        = {Xu, Zhangchen and Jiang, Fengqing and Niu, Luyao and Deng, Yuntian and Poovendran, Radha and Choi, Yejin and Lin, Bill Yuchen},
  booktitle     = {International Conference on Learning Representations},
  pages         = {76346--76382},
  title         = {{Magpie}: Alignment Data Synthesis from Scratch by Prompting Aligned LLMs with Nothing},
  url           = {https://proceedings.iclr.cc/paper_files/paper/2025/hash/be06e3802e9411381feece79b4d960c1-Abstract-Conference.html},
  year          = {2025}
}

@misc{midtool,
  title         = {{MidTool}: Mid-training Data Synthesis for Agentic Tool Use},
  author        = {Jiang, Fengqing and Wang, Yite and Liu, Boyi and Wang, Zhaoyang and Xu, Canwen and Yao, Zhewei and Poovendran, Radha and He, Yuxiong},
  year          = {2026},
  eprint        = {2608.20314},
  archivePrefix = {arXiv},
  url           = {https://arxiv.org/abs/2608.20314}
}

@misc{visualsphinx,
  title         = {{VisualSphinx}: Large-Scale Synthetic Vision Logic Puzzles for {RL}},
  author        = {Feng, Yichen and Xu, Zhangchen and Jiang, Fengqing and Li, Yuetai and Ramasubramanian, Bhaskar and Niu, Luyao and Lin, Bill Yuchen and Poovendran, Radha},
  year          = {2025},
  eprint        = {2505.23977},
  archivePrefix = {arXiv},
  url           = {https://arxiv.org/abs/2505.23977}
}

@misc{scaleenv,
  title         = {{ScaleEnv}: Scaling Environment Synthesis from Scratch for Generalist Interactive Tool-Use Agent Training},
  author        = {Tu, Dunwei and Hao, Hongyan and Yang, Hansi and Chen, Yihao and Zhang, Yi-Kai and Xia, Zhikang and Yang, Yu and Sun, Yueqing and Liu, Xingchen and Shen, Furao and Gu, Qi and Su, Hui and Cai, Xunliang},
  year          = {2026},
  eprint        = {2602.06820},
  archivePrefix = {arXiv},
  url           = {https://arxiv.org/abs/2602.06820}
}

@inproceedings{awm,
  title         = {Agent World Model: Infinity Synthetic Environments for Agentic Reinforcement Learning},
  author        = {Wang, Zhaoyang and Xu, Canwen and Liu, Boyi and Wang, Yite and Han, Siwei and Yao, Zhewei and Yao, Huaxiu and He, Yuxiong},
  booktitle     = {Proceedings of the 43rd International Conference on Machine Learning},
  pages         = {131469--131511},
  year          = {2026},
  volume        = {306},
  series        = {Proceedings of Machine Learning Research},
  publisher     = {PMLR},
  url           = {https://proceedings.mlr.press/v306/wang26jh.html}
}

@misc{step37flash,
  title = {{Step-3.7-Flash} Model Card},
  author = {{StepFun}},
  year = {2026},
  howpublished = {\url{https://huggingface.co/stepfun-ai/Step-3.7-Flash}}
}

\appendix
\section{Supplementary \ourwork{} Details}
\label{app:supplementary}

This section provides source-processing, workspace, collection and verification details for Section~\ref{sec:construction}.

\subsection{Text sources and task generation}
\label{app:text-construction}

We collect licensed GitHub repositories created from 2022 onward through keyword searches and manual selection. GPT-5.5 summarizes their functionality and interactions after excluding third-party libraries, minified files and binaries; repositories with insufficient usable content are discarded. We retain source URLs and licenses. Task generation changes the scenario while preserving the seed's interaction mechanism. Detailed-task generation also receives the scaffold description.

\subsection{Visual reference collection and filtering}
\label{app:visual-construction}

\paragraph{Site selection.} We collect pages from July~17 to September~13, 2026, starting from FineWeb~\citep{fineweb} and filtering with the CrUX site list, metadata, accessibility and \texttt{robots.txt} checks. We shuffle eligible sites, remove duplicate registered domains, and keep sites disjoint across task variants.

\paragraph{Screenshots and assets.} We capture full pages after triggering deferred loading, freezing animations and removing advertising and consent overlays. We reject pages taller than 8,000 pixels or missing visible images. Multi-viewport references use $1920\times1080$, $1024\times768$ and $375\times812$ browser contexts, with mobile input settings for the last. Every capture must meet the height limit and stay within 2\% of the requested width. Images, fonts and media are saved with the references.

\paragraph{Human review.} We inspect references for visible defects, inappropriate content and exposed private information. Multi-page and multi-viewport references are reviewed as groups; a problem in any member rejects the group.

\paragraph{Navigation and task inputs.} Multi-page tasks use three to seven pages collected within two links of the homepage, with visible link labels and navigation paths. Fixed templates combine the retained screenshots, assets, navigation information and viewport dimensions into task statements.

\paragraph{Provenance.} We retain source URLs, crawl timestamps, crawler versions, per-page screenshot hashes and asset source records.

\begin{table}[t]
\centering
\small
\setlength{\tabcolsep}{5pt}
\renewcommand{\arraystretch}{1.1}
\begin{tabular*}{\textwidth}{@{\extracolsep{\fill}}lllll@{}}
\toprule
UI stack & Language & State library & Styling & Build tool \\
\midrule
React 18 & TypeScript & \textemdash{} & Tailwind & Vite \\
\midrule
Vue 3 & TypeScript & \textemdash{} & Tailwind & Vite \\
\midrule
React 18 & TypeScript & Jotai & Plain CSS & Vite \\
React 18 & TypeScript & MobX & Plain CSS & Vite \\
React 18 & TypeScript & Redux Toolkit & Plain CSS & Vite \\
React 18 & TypeScript & Zustand & Plain CSS & Vite \\
\midrule
React 18 & TypeScript & \textemdash{} & styled-components & Vite \\
\midrule
Svelte 5 & TypeScript & \textemdash{} & Plain CSS & Vite \\
\midrule
Angular 19 & TypeScript & \textemdash{} & Plain CSS & Angular CLI \\
\midrule
Plain HTML/JS & JavaScript & \textemdash{} & Tailwind (CDN) & None \\
\midrule
Plain HTML/JS & JavaScript & \textemdash{} & Sass & Vite \\
Plain HTML/JS & JavaScript & \textemdash{} & Less & Vite \\
Plain HTML/JS & JavaScript & \textemdash{} & Stylus & Vite \\
\midrule
Plain HTML/JS & JavaScript & \textemdash{} & Plain CSS & webpack 5 \\
Plain HTML/JS & JavaScript & \textemdash{} & Plain CSS & Parcel 2 \\
\bottomrule
\end{tabular*}
\caption{Fifteen project scaffolds for detailed tasks, arranged in nine groups. Each row is one scaffold; \textemdash{} denotes no additional state-management library.}
\label{tab:project-scaffolds}
\end{table}

\subsection{Project scaffolds}
\label{app:project-scaffolds}

Detailed tasks use 15 reusable scaffolds covering UI frameworks, state management, styling and bundlers, as shown in Table~\ref{tab:project-scaffolds}. Each provides entry files, dependency manifests and build configuration; its stack description is also supplied during task generation. The Angular scaffold uses standalone components and includes router and forms. High-level tasks start from an empty application workspace; the solver chooses and installs its stack.

\subsection{The agent harness}
\label{app:scaffold}

Harbor manages containers and workspaces, and OpenCode runs the coding agent. The release pins OpenCode 1.18.30 for solving, with shell, reading, editing, search and todo tools. GUI verification uses OpenCode 1.18.11 and Playwright CLI 0.1.17. The solving configuration disables \texttt{task}, whose delegated actions are absent from the exported trajectory, and \texttt{webfetch}, to restrict live-page retrieval.

The dataset includes trajectories collected with diverse tool configurations and system prompts. We reconstruct each sample's actual system prompt, tool definitions and user inputs from recorded requests and sessions, including the reference images attached to the first message.

\begin{table}[t]
\centering
\footnotesize
\renewcommand{\arraystretch}{1.15}
\setlength{\tabcolsep}{3pt}
\begin{tabular}{@{}lL{.52\linewidth}@{}}
\toprule
Path & Contents \\
\midrule
\texttt{spec.json} & Task and step statements, categories, source records, fixed checks, starting-state information, repository commit and image digest. \\
\texttt{start/} & Saved initial project files, when needed. \\
\texttt{environment/} & Reference screenshots, assets and multi-page navigation information. \\
\texttt{verifier/}\allowbreak\texttt{task.json} & Visual targets, viewports and reference-image mapping. \\
\texttt{solution/<step>/} & Reference implementation for each step. \\
\texttt{manifest.json} & SHA-256 hashes of the package files. \\
\bottomrule
\end{tabular}
\caption{Task-package files. Optional directories are included when required by the task.}
\label{tab:package-fields}
\end{table}

\subsection{Model roles and conversation continuation}
\label{app:batch}

\paragraph{Model roles.} GPT-5.5 extracts text seeds, and GPT-5.4, GPT-5.5 and GPT-5.6 Sol generate text tasks. We collect trajectories with GPT-5.6 Sol, Qwen3.8-27B, GPT-5.5, MiniMax-M3, GPT-5.4 and Qwen3.5-35B-A3B; their shares are reported in Appendix~\ref{app:statistics}. GPT-5.4, GPT-5.5 and GPT-5.6 Sol construct or review browser checks, while GPT-5.6 Sol performs GUI verification and visual grading.

\paragraph{Interruption and continuation.} We interrupt attempts when a loop detector identifies persistent repetition in tool calls, preserving the workspace and session. Before continuation, we shorten consecutive runs of repeated tool-call messages while retaining the rest of the history. The continuing model resumes from this workspace and trimmed session. Only the resumed portion is supervised; the initial agent's history is masked from training loss.

\subsection{Decontamination}
\label{app:decontamination}

For text-grounded tasks, we compare statements with the 1,029 distinct test statements across the three benchmarks. Following WebGen-Bench~\citep{webgenbench}, we flag word 5-gram Jaccard similarity above $0.6$ or all-MiniLM-L6-v2 embedding cosine similarity above $0.55$. For vision-grounded tasks, we match source registered domains against Vision2Web task names using exact matches, substrings of at least six characters, or shared non-generic words of at least five characters.

We manually review all screening hits. Only confirmed false positives remain eligible for training; other flagged samples are excluded. The final release retains 514 text-grounded trajectories flagged by semantic similarity, none by the lexical threshold, and four visual trajectories confirmed as name collisions. These 518 trajectories are annotated with their matched benchmark tasks and grouped separately within the 15K release. Counts refer to trajectories, not distinct task statements or website pairs.

\begin{table}[t]
\centering
\begingroup
\footnotesize
\setlength{\tabcolsep}{4pt}
\renewcommand{\arraystretch}{1.23}
\begin{tabular*}{\textwidth}{@{\extracolsep{\fill}}lrr@{\hspace{14pt}}rr@{}}
\toprule
\multirow{2}{*}{Technology} & \multicolumn{2}{c}{\textbf{Environments: initial workspaces}} & \multicolumn{2}{c}{\textbf{Trajectories: observed code}} \\
\cmidrule(lr){2-3}\cmidrule(l){4-5}
 & Text-grounded & Vision-grounded & Text-grounded & Vision-grounded \\
\midrule
React & 4,154 (39.6\%) & 549 (12.2\%) & 6,859 (65.3\%) & 549 (12.2\%) \\
vanilla & 1,613 (15.4\%) & 1,286 (28.6\%) & 3,014 (28.7\%) & 3,951 (87.8\%) \\
Vue & 527 (5.0\%) & 0 (0.0\%) & 527 (5.0\%) & 0 (0.0\%) \\
Angular & 60 (0.6\%) & 0 (0.0\%) & 60 (0.6\%) & 0 (0.0\%) \\
Svelte & 40 (0.4\%) & 0 (0.0\%) & 40 (0.4\%) & 0 (0.0\%) \\
Empty app workspace & 4,106 (39.1\%) & 2,665 (59.2\%) & N/A & N/A \\
\midrule
\textbf{Total} & \textbf{10,500 (100\%)} & \textbf{4,500 (100\%)} & \textbf{10,500 (100\%)} & \textbf{4,500 (100\%)} \\
\bottomrule
\end{tabular*}
\par\vspace{4pt}
\begin{minipage}{\textwidth}
\footnotesize\raggedright
Cells show counts (column percentages), rounded independently. Environments are counted per trajectory entry and describe initial workspaces; trajectory classifications use framework imports in read or written code and edits to JSX/TSX files. vanilla denotes plain HTML/CSS/JS. N/A means not applicable: an empty app workspace describes an initial state, not a technology category for trajectory code.\par
\end{minipage}
\endgroup
\caption{Technologies in environment initial workspaces and code observed in trajectories: 15K \ourenv{} entries and 15K \ourdata{} trajectories.}
\label{tab:stacks}
\end{table}

\begin{table}[t]
\centering
\footnotesize
\begin{tabular}{lrrrrrr}
\toprule
Family & $\leq$10 & 11--20 & 21--50 & 51--100 & $>$100 & Total \\
\midrule
Text-grounded  & 496 & 4,786 & 5,038 & 176 & 4 & 10,500 \\
Vision-grounded & 0 & 4 & 860 & 3,299 & 337 & 4,500 \\
\midrule
Total & 496 & 4,790 & 5,898 & 3,475 & 341 & 15K \\
\bottomrule
\end{tabular}
\caption{Distribution of tool calls per trajectory.}
\label{tab:difficulty}
\end{table}

\subsection{Task packaging and validation}
\label{app:env}

Each trajectory has a corresponding environment entry in \ourenv{}. Table~\ref{tab:package-fields} shows packages preserve the statement, starting workspace, reference materials, fixed checks and reference solution. Later iterative tasks start from the preceding solution, handovers from a failed implementation, and conversation continuations from the interrupted workspace and dialogue.

Verification records preserve browser-test code and outcomes, GUI-agent interactions and verdicts, or rendered screenshots and visual assessments, depending on the verifier. Execution failures or incomplete grading records trigger a limited number of retries.

The release tools convert packages to executable Harbor tasks and replay saved solutions against the fixed verifiers. All tested reference solutions pass in the reconstructed environments. The digest-pinned runtime includes Python 3.12, Node.js 22, Chrome, Playwright 1.49.0 and scaffold dependencies. Reference solutions are omitted from the agent's solving workspace.

\section{Data Details}
\label{app:data}

\subsection{Coverage and trajectory statistics}
\label{app:statistics}

Table~\ref{tab:stacks} distinguishes initial-workspace technologies from frameworks observed in trajectory code. Framework counts include existing scaffold code. The median trajectory length is 26.8K tokens, with P90 66.9K, P99 91.5K and a maximum of 98.3K. Tool calls have a median of 26 and P90 of 79; Table~\ref{tab:difficulty} gives their distribution. Aggregate totals and task-family medians appear in Section~\ref{sec:composition}. Table~\ref{tab:text-axes} breaks down text-grounded trajectories and environments by specification level and development setting.

Token counts use the Qwen3.5-35B-A3B processor and ms-swift training template, including messages, tool definitions and visual tokens but excluding the top-level system prompt. Retained continuation history contributes to token and tool-call counts despite being masked from loss. Each tool call counts once, and images are counted by occurrences, including repeated inputs.

Retained trajectories are contributed by multiple members of the trajectory model pool: GPT-5.6 Sol accounts for 75.3\%, Qwen3.8-27B for 8.4\%, GPT-5.5 for 6.9\%, MiniMax-M3 for 4.3\%, and GPT-5.4 for 4.2\%. The remaining 1.0\% comes from Qwen3.5-35B-A3B initial attempts that passed directly. Percentages are rounded independently.

\begin{table}[t]
\centering
\small
\setlength{\tabcolsep}{4pt}
\begin{tabular}{@{}llr@{}}
\toprule
Specification & Setting & Count \\
\midrule
High-level & Standalone & 4,500 \\
High-level & Iterative & 1,000 \\
Detailed & Standalone & 1,855 \\
Detailed & Iterative & 3,145 \\
\midrule
\textbf{Total} & & \textbf{10,500} \\
\bottomrule
\end{tabular}
\caption{Text-grounded trajectories and environments by specification level and development setting. Counts are identical for both.}
\label{tab:text-axes}
\end{table}

\subsection{Verifier audit}
\label{app:audit}

One author audits 600 task-solving attempts sampled before final data selection: 100 accepted and 100 rejected attempts per verifier. Each audit inspects the task statement, submitted application and verification record. Table~\ref{tab:audit} reports accuracy, precision, recall and F1 against the human judgments, treating acceptance as the positive class.

\begin{table}[h]
\centering
\renewcommand{\arraystretch}{1.15}
\footnotesize
\setlength{\tabcolsep}{3pt}
\begin{tabular}{@{}lrrrr@{}}
\toprule
Verifier & Accuracy & Precision & Recall & F1 \\
\midrule
Browser scripts & 94.5 & 98.0 & 91.6 & 94.7 \\
GUI agent & 95.5 & 97.0 & 94.2 & 95.6 \\
Visual judge & 85.0 & 87.0 & 83.7 & 85.3 \\
\bottomrule
\end{tabular}
\caption{Verifier performance on 600 manually audited attempts. Values are percentages. Each verifier is evaluated on 100 originally accepted and 100 originally rejected attempts, with human judgments as the reference.}
\label{tab:audit}
\end{table}

TP and TN denote verifier decisions that agree with human acceptance and rejection, respectively; FP and FN denote incorrect acceptance and rejection. Accuracy is $(\mathrm{TP}+\mathrm{TN})/N$, precision is $\mathrm{TP}/(\mathrm{TP}+\mathrm{FP})$, recall is $\mathrm{TP}/(\mathrm{TP}+\mathrm{FN})$, and F1 is $2\mathrm{TP}/(2\mathrm{TP}+\mathrm{FP}+\mathrm{FN})$. These metrics describe the audit sample, which is balanced by the verifier's original decision.

\section{Data Example}
\label{app:examples}

\begingroup
\def\DataExampleHeading{}
We illustrate the data in \ourdata{} with a standalone development task.

\providecommand{\DataExampleHeading}{\section{Data Example}}
\DataExampleHeading
\label{app:development-case}
\begingroup
\newcommand{\DataExampleCodeSize}{\linespread{1}\fontsize{10pt}{11pt}\selectfont}
\paragraph{PlotPatch: an allotment planner.}
The task asks for an interactive garden planner in which users arrange crops
and paths on a grid, track crop areas, and save and reopen named plans.
The agent finds an empty workspace and chooses React and Vite to implement
PlotPatch (Figure~\ref{fig:plotpatch-task}). Cell painting updates one grid cell,
while Bed Fill traverses adjacent cells with the same original material.
Crop-area counts are recomputed from the updated grid. Named plans are stored
locally and reopened by restoring the saved grid.
The first production build succeeds, but type checking reports missing React
declarations and lint flags component exports and generated identifiers. The
agent adds the declarations, exports the components, and replaces the identifier
generation with a counter. It reruns the build, type check, and lint, then starts
the development server and receives an HTTP 200 response from localhost.

\definecolor{DataExampleUser}{RGB}{246,240,250}
\definecolor{DataExampleModel}{RGB}{244,248,253}
\definecolor{DataExampleToolCall}{RGB}{224,236,249}
\definecolor{DataExampleToolResponse}{RGB}{237,246,238}
\newtcolorbox{DataExampleMessage}[2]{enhanced jigsaw,breakable,
  colback=#1,colframe=black!20,colbacktitle=#1,coltitle=black,
  fonttitle=\bfseries,title={#2},boxrule=.4pt,arc=1mm,
  left=2.5mm,right=2.5mm,top=1mm,bottom=1mm,
  before skip=4pt,after skip=4pt,before upper={\setlength{\parskip}{0pt}}}
\captionsetup{type=figure}
\makeatletter
\@beginparpenalty=10000\relax
\makeatother
\fvset{fontsize=\DataExampleCodeSize,breaklines=true,breakanywhere=true,
  breakautoindent=false,breaksymbolleft={},breaksymbolright={},
  listparameters={\setlength{\topsep}{2pt}\setlength{\partopsep}{0pt}},
  frame=none,xleftmargin=0pt,formatcom=\setlength{\parskip}{0pt}}
\begin{DataExampleMessage}{DataExampleUser}{User input}
\normalsize
PlotPatch is a single-surface allotment planner for gardeners arranging crops and paths on a square plot grid. It opens with a built-in demonstration plan showing mixed planting beds and access routes. The user chooses a crop or ground type from a visual palette, then clicks cells to place it; switching to Bed Fill changes a connected area of matching cells, while Clear Cell removes individual assignments. The selected tool and palette choice remain visibly highlighted, and a live summary updates the occupied area for each crop type. Clear Plot empties the grid after confirmation, while Suggest Layout replaces it with a varied planting arrangement. Users can save the current design under a typed name, see it appear in a saved-plans list, and reopen any saved plan to restore its grid. Give the workspace a practical garden-journal feel with earthy colors, crisp cell boundaries, and controls grouped beside the plot.
\end{DataExampleMessage}

\par\medskip\Needspace{8\baselineskip}\noindent\textbf{Agent trajectory (selected steps)}\par\nobreak

\par\medskip\Needspace{5\baselineskip}\noindent\textbf{Step 3: Inspect the existing workspace}\par\nobreak
\begin{DataExampleMessage}{DataExampleToolCall}{Tool call: \texttt{read}}
\begin{Verbatim}
filePath: /app
\end{Verbatim}
\end{DataExampleMessage}
\begin{DataExampleMessage}{DataExampleToolResponse}{Tool response}
\begin{Verbatim}
<path>/app</path>
<type>directory</type>
<entries>
.git/

(1 entries)
</entries>
\end{Verbatim}
\end{DataExampleMessage}

\par\medskip\Needspace{5\baselineskip}\noindent\textbf{Step 4: Check the installed tooling}\par\nobreak
\begin{DataExampleMessage}{DataExampleModel}{Model response}
\begin{Verbatim}
I’m checking the installed Node tooling, then I’ll scaffold the app in the empty repository.
\end{Verbatim}
\end{DataExampleMessage}

\par\medskip\Needspace{5\baselineskip}\noindent\textbf{Step 7: Implement grid editing and live crop counts}\par\nobreak
\begin{DataExampleMessage}{DataExampleToolCall}{Tool call: \texttt{write}}
\begin{Verbatim}
filePath: /app/src/main.jsx
content (excerpts):
[...]
  const counts = useMemo(() => {
    const next = Object.fromEntries(palette.map((item) => [item.id, 0]))
    grid.forEach((value) => {
      if (value) next[value] += 1
    })
    return next
  }, [grid])
[...]
  const handleCell = (index) => {
    setGrid((previous) => {
      const next = [...previous]
      if (tool === 'clear') {
        next[index] = null
        return next
      }
      if (tool === 'paint') {
        next[index] = selected
        return next
      }

      const target = previous[index]
      if (target === selected) return previous
      const queue = [index]
      const visited = new Set([index])
      while (queue.length) {
        const current = queue.shift()
        next[current] = selected
        const row = Math.floor(current / SIZE)
        const col = current % SIZE
        const neighbors = []
        if (row > 0) neighbors.push(current - SIZE)
        if (row < SIZE - 1) neighbors.push(current + SIZE)
        if (col > 0) neighbors.push(current - 1)
        if (col < SIZE - 1) neighbors.push(current + 1)
        neighbors.forEach((neighbor) => {
          if (!visited.has(neighbor) && previous[neighbor] === target) {
            visited.add(neighbor)
            queue.push(neighbor)
          }
        })
      }
      return next
    })
  }
[...]
  const openPlan = (plan) => {
    setGrid([...plan.grid])
    setCurrentName(plan.name)
    notify(`Opened “${plan.name}”`)
  }
[...]
\end{Verbatim}
\end{DataExampleMessage}
\begin{DataExampleMessage}{DataExampleToolResponse}{Tool response}
\begin{Verbatim}
Wrote file successfully.
\end{Verbatim}
\end{DataExampleMessage}

\par\medskip\Needspace{5\baselineskip}\noindent\textbf{Step 12: Run build, type, and lint checks}\par\nobreak
\begin{DataExampleMessage}{DataExampleToolCall}{Tool call: \texttt{bash}}
\begin{Verbatim}
command: npm run build
workdir: /app
timeout: 120000
\end{Verbatim}
\end{DataExampleMessage}
\begin{DataExampleMessage}{DataExampleToolCall}{Tool call: \texttt{bash}}
\begin{Verbatim}
command: npm run typecheck
workdir: /app
timeout: 120000
\end{Verbatim}
\end{DataExampleMessage}
\begin{DataExampleMessage}{DataExampleToolCall}{Tool call: \texttt{bash}}
\begin{Verbatim}
command: npm run lint
workdir: /app
timeout: 120000
\end{Verbatim}
\end{DataExampleMessage}
\begin{DataExampleMessage}{DataExampleToolResponse}{Tool responses (batch excerpt)}
\begin{Verbatim}
node_modules/lucide-react/dist/lucide-react.d.ts(1,24): error TS7016: Could not find a declaration file for module 'react'. '/app/node_modules/react/index.js' implicitly has an 'any' type.
\end{Verbatim}
\end{DataExampleMessage}

\par\medskip\Needspace{5\baselineskip}\noindent\textbf{Step 13: Plan fixes for the reported errors}\par\nobreak
\begin{DataExampleMessage}{DataExampleModel}{Model response}
\begin{Verbatim}
Lint found entry-point export and generated-ID issues; type checking also needs React declarations. I’m applying those fixes before rerunning checks.
\end{Verbatim}
\end{DataExampleMessage}

\par\medskip\Needspace{5\baselineskip}\noindent\textbf{Step 21: Install the missing React declarations}\par\nobreak
\begin{DataExampleMessage}{DataExampleToolCall}{Tool call: \texttt{bash}}
\begin{Verbatim}
command: npm install --save-dev @types/react @types/react-dom
workdir: /app
timeout: 120000
\end{Verbatim}
\end{DataExampleMessage}
\begin{DataExampleMessage}{DataExampleToolResponse}{Tool response (excerpt)}
\begin{Verbatim}
added 5 packages, and audited 141 packages in 1s
\end{Verbatim}
\end{DataExampleMessage}

\par\medskip\Needspace{5\baselineskip}\noindent\textbf{Step 22: Rerun build, type, and lint checks}\par\nobreak
\begin{DataExampleMessage}{DataExampleToolCall}{Tool call: \texttt{bash}}
\begin{Verbatim}
command: npm run build
workdir: /app
timeout: 120000
\end{Verbatim}
\end{DataExampleMessage}
\begin{DataExampleMessage}{DataExampleToolCall}{Tool call: \texttt{bash}}
\begin{Verbatim}
command: npm run typecheck
workdir: /app
timeout: 120000
\end{Verbatim}
\end{DataExampleMessage}
\begin{DataExampleMessage}{DataExampleToolCall}{Tool call: \texttt{bash}}
\begin{Verbatim}
command: npm run lint
workdir: /app
timeout: 120000
\end{Verbatim}
\end{DataExampleMessage}
\begin{DataExampleMessage}{DataExampleToolResponse}{Tool responses (batch excerpt)}
\begin{Verbatim}
✓ built in 690ms



> plotpatch@1.0.0 typecheck
> tsc --noEmit




> plotpatch@1.0.0 lint
> eslint .
\end{Verbatim}
\end{DataExampleMessage}

\par\medskip\Needspace{5\baselineskip}\noindent\textbf{Step 23: Serve the app and check its response}\par\nobreak
\begin{DataExampleMessage}{DataExampleModel}{Model response}
\begin{Verbatim}
All checks pass. I’m starting the dev server briefly and requesting its localhost page to verify it serves correctly.
\end{Verbatim}
\end{DataExampleMessage}
\begin{DataExampleMessage}{DataExampleToolCall}{Tool call: \texttt{bash}}
\begin{Verbatim}
command: npm run dev > /tmp/opencode/plotpatch-vite.log 2>&1 & server_pid=$!; sleep 2; curl --fail --silent --show-error --head http://localhost:5173/; result=$?; kill "$server_pid"; wait "$server_pid" 2>/dev/null; exit "$result"
workdir: /app
timeout: 120000
\end{Verbatim}
\end{DataExampleMessage}
\begin{DataExampleMessage}{DataExampleToolResponse}{Tool response (excerpt)}
\begin{Verbatim}
HTTP/1.1 200 OK
\end{Verbatim}
\end{DataExampleMessage}

\par\medskip\Needspace{5\baselineskip}\noindent\textbf{Step 26: Report completion}\par\nobreak
\begin{DataExampleMessage}{DataExampleModel}{Model response (final)}
\begin{Verbatim}
Implemented PlotPatch with all requested interactions and responsive styling.

Checks passed: lint, typecheck, build.  
Run `npm run dev` → http://localhost:5173/
\end{Verbatim}
\end{DataExampleMessage}

\caption{PlotPatch task and selected agent steps.}
\label{fig:plotpatch-task}

\endgroup

\endgroup

\section{Training and Evaluation Details}
\label{app:train}

\subsection{Training hyperparameters}
\label{app:training-hyperparameters}

All fine-tuning runs in Section~\ref{sec:experiments} use the following configuration, differing only in their training data.

\begin{center}
\footnotesize
\renewcommand{\arraystretch}{1.15}
\setlength{\tabcolsep}{3pt}
\begin{tabular}{@{}L{.37\columnwidth}L{.58\columnwidth}@{}}
\toprule
Original model & Qwen3.5-35B-A3B \\
Framework & Megatron-LM, through ms-swift \\
Hardware & 8 NVIDIA H100 NVL GPUs \\
Parallelism & tensor 2, context 4, pipeline 1, \\
            & expert 4, expert-tensor 2 \\
            & sequence parallelism enabled \\
\midrule
Tuning & all parameters \\
Frozen & vision tower \\
Loss & assistant turns only \\
Optimizer & Adam with precision-aware updates; master weights fp32, gradients and both moments bf16 \\
\midrule
Adam $\beta_1$, $\beta_2$ / $\epsilon$ & $0.9$, $0.95$ / $10^{-8}$ \\
Weight decay / gradient clipping & $0.1$ / $1.0$ \\
Training / data seed & 42 / 42 \\
Epochs & 2 \\
Global / micro batch & 16 / 1 \\
Learning rate & $2\times10^{-5}$ \\
Schedule / warmup & cosine to $0$ / $10\%$ \\
Length limit & 106{,}496 tokens \\
Sequence packing & on, at 98{,}304 tokens \\
Validation split & $3\%$ \\
\bottomrule
\end{tabular}
\end{center}

\subsection{StepFun-Mix construction}
\label{app:stepfun-data}

StepFun-Mix (15K) is a training subset we assemble from the public Step-3.5-Flash-SFT corpus~\citep{stepfun}, using its \texttt{json/general/} data. It contains 13,500 front-end webpage-generation examples and 1,500 multi-turn software-engineering trajectories from the same corpus. The latter add examples of tool use and execution feedback. They account for 10\% of samples and 37.9\% of tokens.

For webpage generation, we retain instruction-response pairs whose main output is a complete single-file HTML document of at least 2,000 characters, remove mismatched pairs, and deduplicate instructions and responses. Software-engineering trajectories are identified by tool-result turns together with OpenHands prompts or workspace-path markers. Both components undergo benchmark-overlap screening before mixing. We convert them to the training format while retaining reasoning content and tool interactions, deduplicate the combined data, and verify that all 15K samples encode within the 98,304-token packing limit.

\begin{table}[t]
\centering
\footnotesize
\setlength{\tabcolsep}{3pt}
\renewcommand{\arraystretch}{1.15}
\begin{tabular}{@{}lccc@{}}
\toprule
Model & Visual & Functional & Composite \\
\midrule
\modelicon{\includegraphics[width=.9em,height=.9em,keepaspectratio]{figures/model-icons/qwen-color.pdf}} Qwen3.5-35B-A3B & 32.7 & 48.3 & 40.3 \\
\rowcolor[RGB]{235,250,235}
\quad \textbf{+ \ourdata{}} & \textbf{37.6} & \textbf{48.5} & \textbf{42.5} \\
\bottomrule
\end{tabular}
\caption{Supplementary Vision2Web Level~2 results. The + row fine-tunes Qwen3.5-35B-A3B on \ourdata{}. Composite averages the visual and functional scores within each run; each column reports the median across three runs. Higher is better; bold marks the higher score in each column.}
\label{tab:v2w-l2}
\end{table}

\subsection{Evaluation setup}
\label{app:eval-setup}

\paragraph{Harness and model access.} We implement OpenCode adapters for the three benchmarks. The adapters expose the benchmark's task statement and workspace to the agent and return the resulting application to its evaluator. Models retain the tools and system prompt selected by their OpenCode configuration. We use the recommended inference settings for each evaluated model.

\paragraph{Judges and aggregation.} Vision2Web visual grading uses GPT-5; its supplementary Level~2 functional tests use Claude Sonnet 4.5. WebGen-Bench uses GPT-5.4. Web-Bench uses the benchmark's test scripts without an LLM judge. The same judge is used for all compared models within each benchmark.

\begin{itemize}
\setlength\itemsep{3pt}\setlength\parsep{0pt}
\item \textbf{Vision2Web Level 1.} We evaluate 100 tasks at desktop, tablet and mobile viewports. Each viewport's score averages the judge's region scores; the reported average aggregates all 300 viewport scores. Missing evaluations contribute zero. Visual scoring uses the benchmark's reference-comparison procedure, run after screenshots are captured.
\item \textbf{WebGen-Bench.} We evaluate 101 applications. Checks receive 1 for YES, $0.5$ for PARTIAL and 0 for NO or missing results. Overall accuracy pools the checks across applications, rather than averaging application-level rewards. The functional, data-display and design-validation columns apply the same calculation within each check category. Checks from applications without usable verifier outputs remain in the denominator with zero credit. We do not evaluate the separate appearance-score metric.
\item \textbf{Web-Bench.} We evaluate 50 projects with 20 sequential tasks each, reporting pass@1 and pass@2 with the official scoring procedure. Task progression depends on success at earlier steps. Consequently, scores describe sequential development performance, and failure at an earlier step can prevent later tasks from being reached.
\end{itemize}

These results use our OpenCode integrations and fixed judges; comparisons with published leaderboards require accounting for these setup differences.

\subsection{Vision2Web Level 2 results}
\label{app:v2w-l2}

Vision2Web Level~2 requires building multi-page interactive frontends from visual references and textual specifications. Due to the high cost of long-horizon evaluation, we compare Qwen3.5-35B-A3B and \ourmodel{}, on its 66 tasks. We evaluate each model three times and report the median of each metric in Table~\ref{tab:v2w-l2}. The composite score is the mean of the visual score (VS) and functional score (FS) within each run.

The median composite score increases modestly from 40.3 to 42.5. The improvement is concentrated in visual fidelity, with median VS rising from 32.7 to 37.6, while median FS remains similar (48.3 versus 48.5). Completing complex multi-page interactions remains a challenge for the fine-tuned model.

\section{Before and After Fine-Tuning}
\label{app:qualitative}

\begin{figure}[t]
\centering
\includegraphics[width=\textwidth]{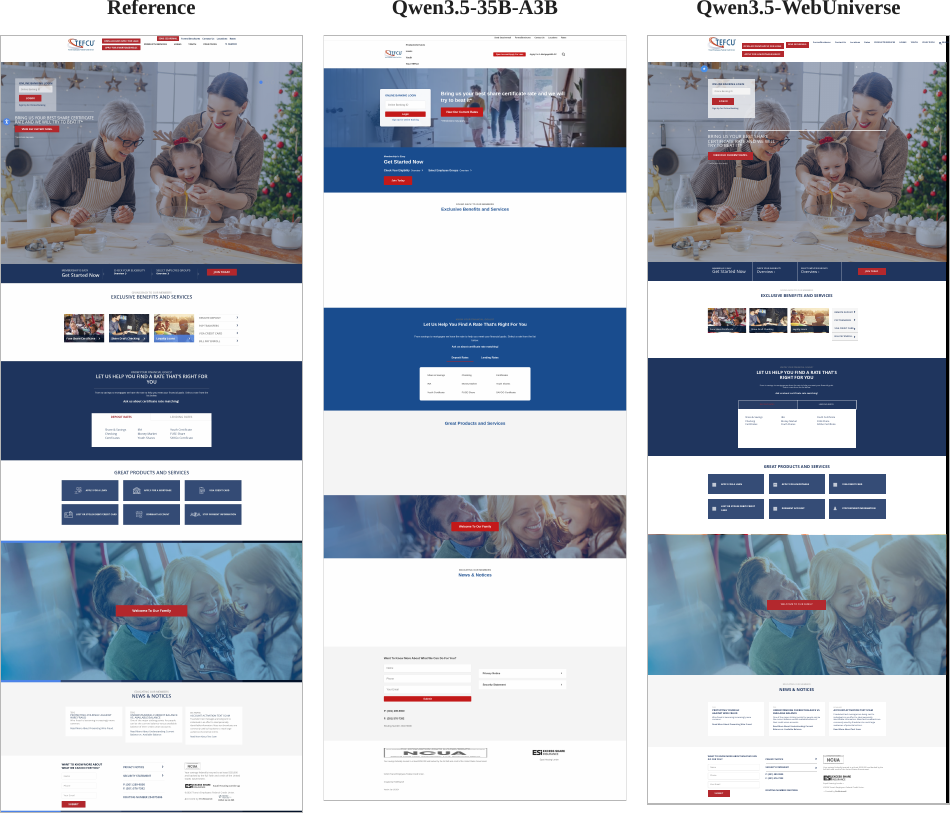}
\caption{Complete desktop pages for the tefcu example in Vision2Web Level~1. The reference and outputs from Qwen3.5-35B-A3B and \ourmodel{} are shown with their aspect ratios preserved.}
\label{fig:qualitative-v2w-full}
\end{figure}

Figure~\ref{fig:qualitative-v2w-full} shows the complete pages for the visual reproduction example in Figure~\ref{fig:qualitative-cases}, comparing Qwen3.5-35B-A3B with \ourmodel{}.

\paragraph{Visual reproduction.} Figure~\ref{fig:qualitative-v2w-full} shows the full tefcu pages. The fine-tuned output restores sections largely absent from the original model's page and more closely matches the reference layout.

\section{Additional Analysis}
\label{app:analysis}

\subsection{Agent behavior before and after fine-tuning}
\label{app:behavior}

We analyze evaluation traces to characterize application inspection, verification and revision before and after fine-tuning. Table~\ref{tab:agent-behavior} compares six behavior metrics for Qwen3.5-35B-A3B and \ourmodel{} across the three benchmarks.

\paragraph{Analysis protocol.} The original-model traces correspond to Table~\ref{tab:main}; the fine-tuned traces come from supplementary runs. We combine retries for each Vision2Web Level~1 and WebGen-Bench task, yielding 100 and 101 task-level units per model. For Web-Bench, we analyze individual attempts: 275 for the original model and 305 for \ourmodel{}. These attempts cover different stages of the sequential benchmark and are not paired. We match 261 and 297 attempts, respectively, to test outcomes; unmatched attempts contribute to behavior frequencies but not outcome comparisons. Each frequency is the percentage of analyzed units with at least one occurrence of the behavior. We identify occurrences from tool names, argument summaries and outputs; metrics may overlap.

\begin{table}[t]
\centering
\footnotesize
\setlength{\tabcolsep}{5pt}
\renewcommand{\arraystretch}{1.15}
\begin{tabular*}{\textwidth}{@{\extracolsep{\fill}}lcccccc@{}}
\toprule
\multirow{2}{*}{Behavior metric} & \multicolumn{2}{c}{Vision2Web Level~1} & \multicolumn{2}{c}{WebGen-Bench} & \multicolumn{2}{c}{Web-Bench} \\
\cmidrule(lr){2-3}\cmidrule(lr){4-5}\cmidrule(l){6-7}
 & Base & Tuned & Base & Tuned & Base & Tuned \\
\midrule
Evaluated units ($n$) & 100 & 100 & 101 & 101 & 275 & 305 \\
\midrule
Pre-edit inspection & 5.0 & 77.0 & 1.0 & 0.0 & 73.8 & 85.2 \\
Post-edit verification & 90.0 & 100.0 & 50.5 & 95.0 & 72.7 & 75.7 \\
Error-triggered revision & 0.0 & 1.0 & 45.5 & 73.3 & 1.1 & 2.0 \\
\midrule
Server-launch timeout & 98.0 & 43.0 & 51.5 & 27.7 & 0.0 & 1.0 \\
Repeated tool calls & 0.0 & 6.0 & 0.0 & 0.0 & 0.0 & 2.0 \\
No implementation edits & 2.0 & 0.0 & 0.0 & 1.0 & 0.7 & 8.5 \\
\bottomrule
\end{tabular*}
\caption{Behavior occurrence rates (\%) in evaluation traces. Base: Qwen3.5-35B-A3B; Tuned: \ourmodel{}. Vision2Web and WebGen-Bench use task-level units with retries combined; Web-Bench uses individual attempts. Tuned-model traces are from supplementary runs. Metrics are defined in the text.}
\label{tab:agent-behavior}
\end{table}

\paragraph{Metric definitions.} \emph{Pre-edit inspection} requires reading file contents or accessing an image before the first implementation edit; directory listings are excluded. \emph{Post-edit verification} requires a build, type check, test, local HTTP request, script or browser check without a detected error after the final edit, or rereading a modified file. \emph{Error-triggered revision} requires a failed check, an implementation edit and a subsequent check of the same kind; the subsequent check need not pass.

\emph{Server-launch timeout} denotes a server-launch tool call that times out without explicit process detachment through \texttt{nohup}, \texttt{setsid} or \texttt{disown}. \emph{Repeated tool calls} requires at least 30 consecutive calls with the same tool-and-argument signature, or 30 repetitions of a two-call pattern. \emph{No implementation edits} denotes an attempt without a detected implementation-file write or edit; changes to temporary files and documentation are excluded.

\paragraph{Pre-edit inspection increases on visual and iterative tasks.} Inspection rates rise from 5.0\% to 77.0\% on Vision2Web and from 73.8\% to 85.2\% on Web-Bench, while remaining low on WebGen-Bench (1.0\% and 0.0\%). Web-Bench Fastify task~2 illustrates the role of existing-code inspection: the original model writes an EJS layout but fails to register it with the view engine, whereas \ourmodel{} inspects the application, connects the layout and passes the task.

\paragraph{Verification and error-triggered revision become more frequent.} Post-edit verification increases across all three benchmarks, including from 50.5\% to 95.0\% on WebGen-Bench. On the same benchmark, error-triggered revision rises from 45.5\% to 73.3\%. In a separate run of task~000020 from Figure~\ref{fig:qualitative-cases}, \ourmodel{} repairs reported TypeScript errors and rebuilds successfully before checking the service; the original model finishes after starting the server. On Web-Bench, the instructions state that tests are unavailable and request code rereading. Most recorded post-edit verification there consists of rereading; execution checks occur in 2.5\% and 4.3\% of attempts, respectively.

\paragraph{Repeated tool calls persist.} Server-launch timeouts decrease on Vision2Web and WebGen-Bench, but repeated calls occur in six fine-tuned Vision2Web traces and six Web-Bench attempts, compared with none for the original model. Examples include 1,311 identical image-dimension queries on liebherr and 604 searches for unavailable tests on Web-Bench Vue task~6. All six flagged Web-Bench attempts fail.

\paragraph{Retries do not consistently revise the implementation.} Among Web-Bench second attempts, 23/72 (31.9\%) for \ourmodel{} contain no recorded implementation edits, compared with 2/83 (2.4\%) for Qwen3.5-35B-A3B. These rates use second attempts as the denominator; Table~\ref{tab:agent-behavior} reports frequencies over all attempts. All 23 no-edit second attempts by \ourmodel{} fail. In Mobx task~1, the agent rereads its files and finishes without modifying the implementation after feedback reports missing required page elements.

\begin{table}[t]
\centering
\footnotesize
\setlength{\tabcolsep}{5pt}
\renewcommand{\arraystretch}{1.15}
\begin{tabular}{@{}lcc@{}}
\toprule
Region score & Base & Tuned \\
\midrule
0 & 12.0 & 2.9 \\
0.25 & 16.1 & 5.2 \\
0.5 & 27.3 & 22.6 \\
0.75 & 40.8 & 50.5 \\
1 & 3.8 & 18.8 \\
\midrule
Evaluated regions ($n$) & 2,709 & 2,739 \\
\bottomrule
\end{tabular}
\caption{Vision2Web Level~1 region-score distributions (\%). Base: Qwen3.5-35B-A3B; Tuned: \ourmodel{}. Percentages are computed within each model's evaluated regions. The tuned-model run is supplementary to Table~\ref{tab:main}.}
\label{tab:region-scores}
\end{table}

\subsection{Vision2Web Level 1 visual analysis}
\label{app:visual-diagnostics}

We examine visual fidelity using the Level~1 traces analyzed in Appendix~\ref{app:behavior}. Table~\ref{tab:region-scores} pools region assessments across desktop, tablet and mobile viewports. The original and fine-tuned outputs contain different numbers of evaluated regions; percentages use each model's region count as the denominator.

\paragraph{Fine-tuning shifts regional scores toward higher fidelity.} The share of regions scored at 0 or 0.25 falls from 28.1\% to 8.1\%, while the share scored at 1 rises from 3.8\% to 18.8\%. Higher aggregate scores in Table~\ref{tab:main} are thus accompanied by fewer low-scoring regions.

\paragraph{Rendering and responsive-layout errors persist.} In the fine-tuned liebherr output, lower-page sections use \texttt{opacity: 0} until scrolled into view and remain hidden in the evaluator's full-page capture, receiving zero on five of seven desktop regions. The agent checks HTTP responses and asset paths without inspecting a rendered screenshot. Mobile-width mismatches also remain: captures exceed the reference width by more than 2\% in 16/100 fine-tuned outputs versus 12/100 original outputs. For rottentomatoes, wide images and navigation yield a 645-pixel capture against a 375-pixel reference, despite a recorded screenshot check.

\subsection{Web-Bench results by technology and task progression}
\label{app:chain-progress}

We compare Qwen3.5-35B-A3B, Qwen3.5-StepFun and \ourmodel{} across Web-Bench's nine technology domains and through its sequential development tasks.

\paragraph{Performance gains vary across technology domains.} \ourmodel{} exceeds Qwen3.5-StepFun in pass@2 on seven domains, ties on full-stack frameworks and scores lower on build tools. Relative to the original model, pass@2 increases from 5.0\% to 25.0\% for UI frameworks and from 23.8\% to 36.2\% for databases, but decreases from 35.0\% to 23.3\% for build tools.

\paragraph{Fine-tuning improves progression through early tasks.} We measure the number of consecutive tasks completed in each of the 50 projects, allowing up to two attempts per task. The fraction completing at least the first four tasks rises from 28\% for the original model to 40\% for Qwen3.5-StepFun and 56\% for \ourmodel{}. Neither fine-tuned model completes all 20 tasks on any project.

\subsection{Vision2Web Level 2 failure analysis}
\label{app:l2-failures}
\begin{table}[!t]
\centering
\footnotesize
\setlength{\tabcolsep}{5pt}
\renewcommand{\arraystretch}{1.15}
\begin{tabularx}{\textwidth}{@{}lXcc@{}}
\toprule
Failure category & Diagnostic criterion & Tasks & Share (\%) \\
\midrule
Browser runtime failure & Blank page during browser evaluation. & 13 & 72.2 \\
Service-port mismatch & Service runs on a port other than the required port. & 4 & 22.2 \\
Execution loop & Repeated tool calls; no launch script delivered before the time limit. & 1 & 5.6 \\
\bottomrule
\end{tabularx}
\caption{Failure categories among 18 tasks with zero visual and functional scores in the inspected \ourmodel{} Vision2Web Level~2 run. Shares use these 18 tasks as the denominator, out of 66 evaluated tasks.}
\label{tab:l2-failures}
\end{table}

We analyze delivery failures in one \ourmodel{} Level~2 run with VS 37.6 and FS 47.3 from Appendix~\ref{app:v2w-l2}. Eighteen of the 66 tasks receive zero on both metrics. Table~\ref{tab:l2-failures} classifies these tasks by observed failure and gives the diagnostic criterion for each category.

\paragraph{Browser runtime failures dominate zero-score tasks.} Thirteen of the 18 tasks produce blank pages. All 13 traces include HTTP checks and ten include build checks, but none records inspection of the agent's rendered screenshot. In holyoke, an unimported icon symbol causes a browser runtime error despite a successful build and HTTP responses. This case shows why build and service checks alone do not establish correct browser execution.

\paragraph{Port mismatches and execution loops prevent delivery.} Four tasks run on the wrong port. In uk\_ixl, the agent observes no response on the required port 3000 and a successful response on 5173, but leaves the launch configuration unchanged. The remaining task, anomali, repeats a directory-creation command 2,459 times and reaches the two-hour limit without delivering a launch script.

\end{document}